\documentclass[a4paper,11pt]{article}
\usepackage{jheppub}

\usepackage{bm}
\usepackage{array}
\usepackage{booktabs}
\usepackage{multirow}
\usepackage{placeins}
\usepackage{xcolor}
\usepackage{orcidlink}

\allowdisplaybreaks[1]

\graphicspath{{output/figures/}}
\DeclareGraphicsExtensions{.pdf,.png}

\newcommand{\GeV}{\mathrm{GeV}}
\newcommand{\ps}{\mathrm{ps}}
\newcommand{\HBM}{\mathrm{HBM}}
\newcommand{\NRQM}{\mathrm{NRQM}}
\newcommand{\Xibcp}{\Xi_{bc}^{+}}
\newcommand{\Xibcz}{\Xi_{bc}^{0}}
\newcommand{\Omegabcz}{\Omega_{bc}^{0}}
\newcommand{\B}{{\cal B}}
\def\lsim{ {\ \lower-1.2pt\vbox{\hbox{\rlap{$<$}\lower5pt\vbox{\hbox{$\sim$}
}}}\ } }
\def\gsim{ {\ \lower-1.2pt\vbox{\hbox{\rlap{$>$}\lower5pt\vbox{\hbox{$\sim$}
}}}\ } }

\title{Mixing-angle sensitivity of inclusive
\texorpdfstring{\(\Xi_{bc}\)}{Xi-bc} and
\texorpdfstring{\(\Omega_{bc}\)}{Omega-bc} lifetimes}

\author[a]{Hai-Yang Cheng\,\orcidlink{0000-0002-4664-697X},}
\author[b]{Chia-Wei Liu\,\orcidlink{0000-0002-3969-6119},}
\author[c]{Fanrong Xu\,\orcidlink{0000-0003-0986-8028}}

\affiliation[a]{Institute of Physics, Academia Sinica, Taipei, Taiwan 11529, Republic of China}
\affiliation[b]{School of Fundamental Physics and Mathematical Sciences,
Hangzhou Institute for Advanced Study, UCAS, Hangzhou 310024, China}
\affiliation[c]{Department of Physics, Jinan University,
Guangzhou 510632, People's Republic of China}

\emailAdd{phcheng@phys.sinica.edu.tw}
\emailAdd{chiaweiliu@ucas.ac.cn}
\emailAdd{fanrongxu@jnu.edu.cn}

\abstract{
	We study the inclusive lifetimes of \(\Xibcp\), \(\Xibcz\), and
	\(\Omegabcz\) with emphasis on the mixing between the spin-zero and
		spin-one \(bc\)-diquark configurations.  The matrix elements are evaluated
	in both the homogeneous bag model (HBM) and the nonrelativistic quark model
	(NRQM).  
	At the benchmark mixing angles \(-23.0^\circ\) for
	\(\Xi_{bc}\) and \(-23.5^\circ\) for \(\Omega_{bc}\), the corresponding
	lifetime triplets are
\[
		(\tau_{\Xibcp},\tau_{\Xibcz},\tau_{\Omegabcz})
		=
		\left(
		{\color{black}0.421^{+0.098}_{-0.093}{}^{+0.007}_{-0.014}},
		{\color{black}0.075^{+0.035}_{-0.017}{}^{+0.004}_{-0.001}},
			{\color{black}0.223^{+0.074}_{-0.046}{}^{+0.018}_{-0.012}}
		\right)\,\ps
\]
	in HBM and
\[
			(\tau_{\Xibcp},\tau_{\Xibcz},\tau_{\Omegabcz})
	= 	\left(
		0.367^{+0.076}_{-0.070}{\color{black}{}^{+0.001}_{-0.005}},
		0.100^{+0.032}_{-0.014}{\color{black}{}^{+0.006}_{-0.001}},
			0.229^{+0.066}_{-0.042}{\color{black}{}^{+0.017}_{-0.011}}
		\right)\,\ps
\]
		in NRQM. {\color{black}The first errors are the fixed-angle uncertainties;
        the second are the envelopes from varying each mixing angle by
        $\pm10^\circ$ with all other inputs fixed.}
        Relative to the unmixed spin-one reference points, the
		\(\Xibcz\) and \(\Omega_{bc}\) lifetimes decrease by approximately
		\(17\%\)--\(19\%\). Neither the minimum nor the maximum of the central lifetime curves occurs
	at a pure \(S_{bc}=0\) or \(S_{bc}=1\) configuration; the extrema instead
	lie at mixed angles with
	\(\lvert\phi_{\cal B}\rvert\simeq27^\circ\)--\(63^\circ\).
	The two pure-spin cases alone are therefore insufficient to characterize
	the angular dependence. 
}

\begin{document}
\maketitle

\section{Introduction}
\label{sec:introduction}

The \(bcq\) baryon system is a distinctive laboratory for the inclusive weak
decay.  Both heavy constituents decay weakly.   
The heavy-quark expansion (HQE) separates the corresponding short-distance
coefficients from local hadronic matrix elements and organizes the
inverse-mass corrections
\cite{Shifman:1984wx,Shifman:1986,Guberina:1986gd,Chay:1990da,
Bigi:1992su,Bigi:1993ex}.  Although the nominal leading term is the free-quark
decay, the flavor-dependent splittings at the leading spectator order arise
from dimension-six four-quark operators.  Their
\(1/m_Q^3\) suppression is compensated by the two-body phase-space factor
\(16\pi^2\), making lifetime hierarchies sensitive to the spin-flavor
wave function and to short-distance quark-pair overlaps
\cite{Neubert:1996we,Lenz:2014jha,Albrecht:2024oyn}.

Ground-state \(bcq\) baryons are among the remaining weakly decaying doubly
heavy baryons yet to be observed.  LHCb's discoveries of
\(\Xi_{cc}^{++}\) \cite{LHCb:2017iph} and, most recently, \(\Xi_{cc}^{+}\) \cite{LHCb:2026xiccp} and $\Omega_{cc}^+$ \cite{YuWang,Xu} together with the
\(\Xi_{cc}^{++}\) lifetime and precision-mass measurements
\cite{LHCb:2018zpl,LHCb:2019epo}, have established
that doubly heavy baryon spectroscopy is experimentally accessible.
Searches for \(\Xi_{bc}^{0}\), \(\Xi_{bc}^{+}\), and
\(\Omega_{bc}^{0}\) have not established any of these states and report
upper limits as functions of the assumed mass and lifetime
\cite{LHCb:2020bc0,LHCb:2021bc0,LHCb:2022bcp}.  Reliable lifetime
expectations are therefore relevant not only to the interpretation of a
future signal, but also to the design and efficiency calibration of the
searches themselves.

For \(bcq\) baryons, the spin composition introduces an additional
ambiguity that is absent when the two heavy quarks are identical.  For an
\(S\)-wave color-antitriplet heavy pair, Fermi statistics requires
\(S_{cc}=S_{bb}=1\), whereas the nonidentical \(bc\) pair allows both
\(S_{bc}=0\) and \(S_{bc}=1\); either can couple with the light quark to
\(J^P=1/2^+\).  Hyperfine interactions mix these two basis states in both
the \(\Xi_{bc}\) and \(\Omega_{bc}\) sectors.  Model estimates span a wide
range: a quark-model calculation finds large mixing, a Bethe--Salpeter
treatment finds percent-level ground-state mixing effects, and a bag-model
spectrum contains sizable admixtures
\cite{Roberts:2007ni,Li:2021qod,Zhang:2021yul}.  The  previous  lifetime analyses   present 
separate predictions for the two pure assignments \(S_{bc}=0\) or \(S_{bc}=1\)~\cite{Cheng:2019sxr,Dulibic:2026}.
The lattice calculation gives
\(\Delta M_\Xi\equiv M_{\Xi_{bc}(S_{bc}=1)}- M_{\Xi_{bc}(S_{bc}=0)}=16(18)_{\rm stat}(38)_{\rm syst}\,\mathrm{MeV}\) 
and
\(\Delta M_\Omega\equiv M_{\Omega_{bc}(S_{bc}=1)}- M_{\Omega_{bc}(S_{bc}=0)}=35(9)_{\rm stat}(25)_{\rm syst}\,\mathrm{MeV}\)
for the two \(J^P=1/2^+\) levels~\cite{Brown:2014ena}.   
Since
\(\sin(2\phi_{\cal B})\propto\Delta M_{\cal B}^{-1}\), with
\(\phi_{\cal B}\) being the mixing angle defined below, a sizable mixing is thus possible.

	\begin{table}[ht]
		\caption{Predicted lifetimes of bottom-charm baryons ${\cal B}_{bc}$ 
			in units of $10^{-13}s$. The quoted predictions from \cite{Dulibic:2026} are for the ${\cal B}_{bc}$ with the axial-vector $bc$-diquark in the kinetic quark mass scheme.
		} \label{tab:lifetimes}
		\begin{center}
			\begin{tabular}{l c c c } \hline \hline
				& ~~$\Xi_{bc}^{+}$~~ & ~~~$\Xi_{bc}^{0}$~~~ & ~~$\Omega_{bc}^{0}$~~ \\
				\hline
				~~Likhoded et al.  ('99) \cite{Likhoded:1999yv}~~ &  $2.8$ & $2.6$  & 2.1  \\
				~~Kiselev et al.  ('99) \cite{Kiselev:1999bc}~~ &  $3.3\pm0.8$ & $2.8\pm0.7$  &  --  \\
		~~Kiselev et al.  ('01) \cite{Kiselev:2001fw}~~ &  $3.0\pm0.4$ & $2.7\pm0.3$  &  $2.2\pm0.4$  \\
				~~Karliner, Rosner ('14)  \cite{Karliner:2014gca}~~ & 2.44 & 0.93 &  -- \\
				~~Berezhnoy et al.  ('18) \cite{Berezhnoy:2018} & $2.4\pm0.2$ & $2.2\pm0.18$ & $1.8\pm0.088$ \\
                ~~Cheng, Xu  ('19) \cite{Cheng:2019sxr} & 4.09\,--\,6.07 & 0.93\,--\,1.18 & 1.68\,--\,3.70 \\
                ~~Dulibi\'c et al. ('26) \cite{Dulibic:2026} & $3.2\pm0.3$ & $0.9\pm0.2$ & $2.0\pm0.3$ \\
				\hline \hline
			\end{tabular}
		\end{center}
	\end{table}

The predicted ${\cal B}_{bc}$ lifetimes available in the literature are collected in Table \ref{tab:lifetimes}. The lifetime hierarchy pattern was found to be $\tau(\Xi_{bc}^+)\gsim\tau(\Xi_{bc}^0)> \tau(\Omega_{bc}^0)$ before 2019. It was modified to $\tau(\Xi_{bc}^+)> \tau(\Omega_{bc}^0) >\tau(\Xi_{bc}^0)$ in 2019 by two of us  (HYC and FRX)~\cite{Cheng:2019sxr} and confirmed lately by the authors in \cite{Dulibic:2026}.   
A $\Xi_{bc}$ width contains charm-quark spectator effects, bottom-quark spectator effects, and heavy-heavy $bc$ spectator effects in the same hadron.
As stressed in \cite{Cheng:2019sxr},  a large constructive $W$-exchange contribution to $\Xi_{bc}^0$  and a large destructive Pauli interference contribution to $\Xi_{bc}^+$ render a substantial lifetime difference between $\Xi_{bc}^+$ and $\Xi_{bc}^0$. 

The $J^P=1/2^+$ ${\cal B}_{bc}$ baryons with $S_{bc}=0$ and $S_{bc}=1$ configurations can mix together
to form a physical state
\begin{equation}
\label{eq:Bbc}
|\B_{bc}(\phi_\B)\rangle = \cos\phi_\B | S_{bc}=1 \rangle+ \sin\phi_\B | S_{bc}=0 \rangle,
\end{equation}
where we have chosen the convention such that the $\B_{bc}$ state at $\phi_\B=0^\circ$
is a pure $S_{bc}=1$ one.  
 For a transition
operator \({\cal T}\), its expectation value in the state of
Eq.~\eqref{eq:Bbc} decomposes as
\begin{equation}
\begin{aligned}
\langle \B_{bc}(\phi_\B)|{\cal T}|\B_{bc}(\phi_\B)\rangle
={}&\frac{{\cal T}_{00}+{\cal T}_{11}}{2}
-\cos(2\phi_\B)\frac{{\cal T}_{00}-{\cal T}_{11}}{2} +\sin(2\phi_\B)\,
\operatorname{Re} ( 
{\cal T}_{01} ) ,
\end{aligned}
\label{eq:Tmixing}
\end{equation}
where \({\cal T}_{ij }\equiv\langle S_{bc} = i |{\cal T}|S_{bc} = j \rangle\).  The endpoint predictions determine
\({\cal T}_{00}\) and \({\cal T}_{11}\), but not \({\cal T}_{01}\).
The pure \(S_{bc}=1\) (\(\phi_\B=0\)) and \(S_{bc}=0\)
(\(\phi_\B=\pi/2\)) limits are stationary if and only if
\(
\operatorname{Re}( 
{\cal T}_{01}) =0\); for \({\cal T}_{00}\neq{\cal T}_{11}\), they are then
the two extrema.

In this work we study the lifetime dependence on the mixing angles
\(\phi_{\Xi}\) and \(\phi_{\Omega}\)~\cite{Yang:2022nps}.    The required four-quark matrix elements are evaluated in two
structurally different descriptions: the homogeneous bag model (HBM), with
relativistic projected bag states, and a Gaussian nonrelativistic quark
model (NRQM), with Jacobi wave functions and contact densities.  
Agreement between the two angular patterns
tests whether the predicted sensitivity survives a substantial change in
the spatial wave function.

The literature on doubly heavy baryons includes lifetime calculations
based on several HQE implementations and nonperturbative inputs, together
with broader studies of their masses, production, decays, and detection
\cite{Kiselev:1998cc,Kiselev:1999bc,Likhoded:1999yv,Guberina:1999,
Chang:2007xa,Kiselev:2001fw,Karliner:2014gca,Cheng:2018mwu,
Cheng:2019sxr,Berezhnoy:2018}.   
Modern HQE studies of singly and doubly charmed baryons and of bottom
baryons have updated the short-distance treatment while exposing the
sensitivity to four-quark matrix elements and higher-dimensional charm
corrections
\cite{Dulibic:2023jeu,Gratrex:2022xpm,Gratrex:2023pfn,Cheng:2023jpz,Cheng:2026mlv}.  
Semileptonic form factors and decay widths of doubly heavy
baryons have also been studied using heavy-quark-symmetry
constraints~\cite{Hernandez:2007qv}.

\begin{figure}[!p]
\centering
\includegraphics[width=0.86\linewidth]{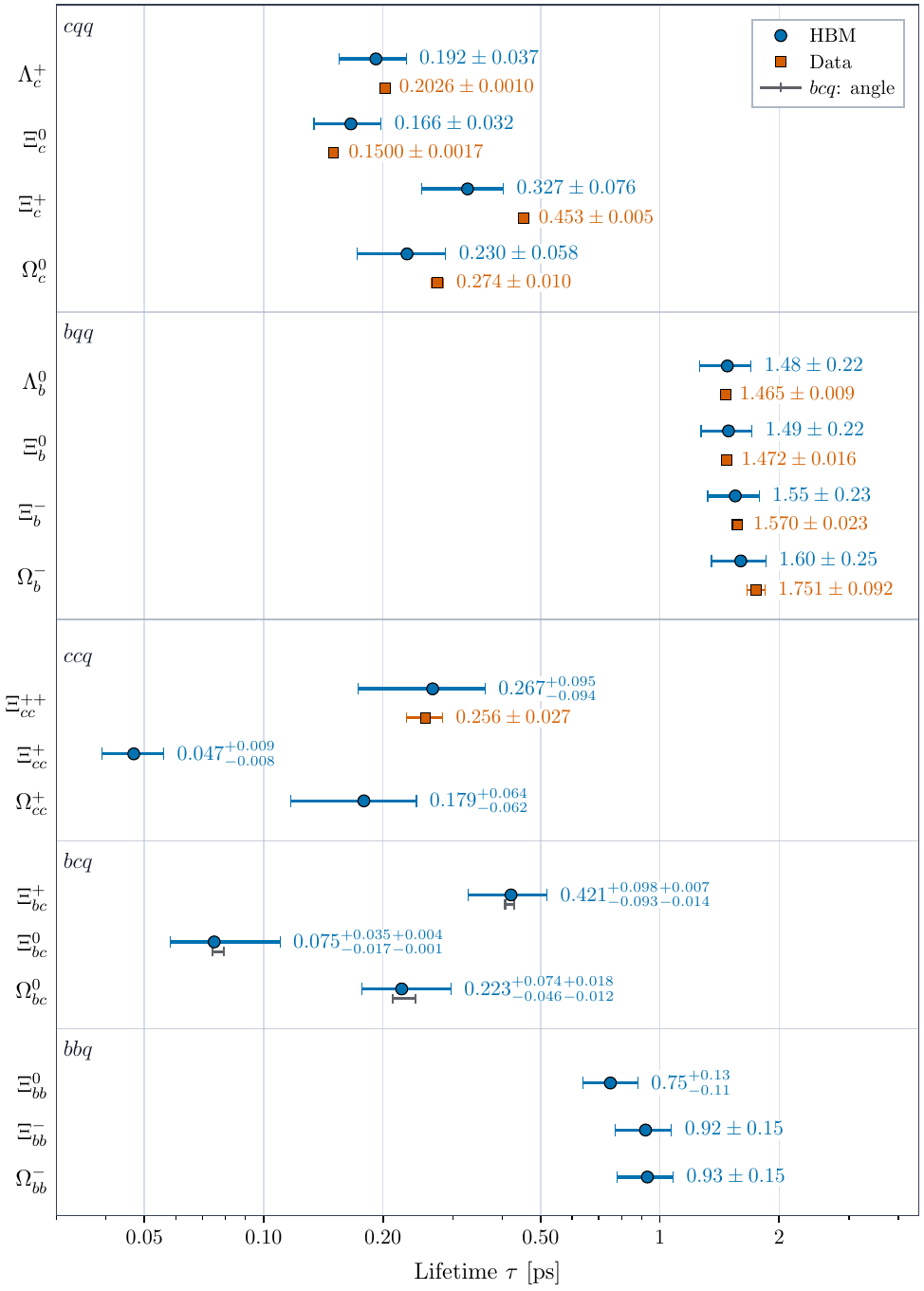}
\caption{HBM lifetime predictions and available experimental data.  The
\(cqq\) and \(bqq\) predictions are from Ref.~\cite{Cheng:2023jpz}, the
\(ccq\) and \(bbq\) predictions from Ref.~\cite{Cheng:2026mlv}, and the
\(bcq\) predictions from this work. The singly heavy experimental
averages are from the PDG~\cite{PDG}, {\color{black}except for
$\Omega_b^-$, for which we use the new LHCb Run~1+Run~2 average~\cite{LHCb:2026OmegaBLifetime},
combining its statistical and systematic errors in quadrature.}
The \(\Xi_{cc}^{++}\) result is from LHCb~\cite{LHCb:2018zpl}, with its
statistical and systematic uncertainties combined and symmetrized for
display. Circles denote HBM predictions and squares denote data.
{\color{black}For $bcq$, the first errors and blue bars are the fixed-angle
uncertainties; the second errors and gray bars, offset downward for visibility,
show the separate $\pm10^\circ$ angle envelopes.}}
\label{fig:hbm-lifetime-atlas}
\end{figure}

In 
fig.~\ref{fig:hbm-lifetime-atlas} we collect the HBM lifetime predictions
across the singly and doubly heavy baryon sectors together with the available
data.  
The HQE, combined with HBM matrix elements, describes the observed pattern of
singly heavy baryon lifetimes surprisingly well; even the charm results remain
compatible with data within their substantially larger theoretical
uncertainties~\cite{Cheng:2023jpz}.
This phenomenological success, however, does not imply uniform order-by-order
convergence.  In \(\Xi_{cc}^{+}\), the dimension-six \(q=d\) $W$-exchange
term represented by the \(cd\to cd\) forward-scattering channel exceeds the
nominal leading contribution, so its absolute lifetime prediction should be
regarded as indicative~\cite{Cheng:2026mlv}.  Dimension-seven spectator
corrections are likewise sizable in channels with a strange spectator: they
are essential for  explaining  the observed placement of \(\Omega_c^0\) in the
singly charmed lifetime hierarchy~\cite{Cheng:2021vca,Cheng:2023jpz},  producing 
substantial cancellations in the $s$-spectator contributions to
\(\Omega_{cc}^{+}\), and partly   canceling the positive dimension-six \(cs\)
contribution in \(\Omega_{bc}^{0}\) found below.
These channels therefore carry enhanced HQE-truncation uncertainty.

This work is organized as follows. In Sec.~\ref{sec:framework} we set up the framework for the weak Hamiltonians, the HQE operator
basis, and the spectator topologies and their signs. We introduce in 
Sec.~\ref{sec:inputs}  two hadronic inputs and the
parametric-sensitivity prescription.  Numerical results and discussions are
presented in Sec.~\ref{sec:numerical}. Conclusions are summarized in
Sec.~\ref{sec:conclusion}.

\section{Theoretical framework}
\label{sec:framework}
 
Both the bottom and charm quarks  decay, and their effective
Hamiltonians are given by~\cite{Buchalla:1995vs}
\begin{align}
 {\cal H}_{\rm eff}^{(b)}
 &=\frac{G_F}{\sqrt2}
 \sum_{\substack{q_3=d,s\\q_1,q_2=u,c}}
 V_{q_1b}^{*}V_{q_2q_3}
 \left(C_1Q_{1,b}^{q_1q_2q_3}
      +C_2Q_{2,b}^{q_1q_2q_3}\right)
 +{\cal H}_{\rm pen}^{(b)}+{\cal H}_{\rm SL}^{(b)}
 +{\rm h.c.},
 \label{eq:heff-b}\\
 {\cal H}_{\rm eff}^{(c)}
 &=\frac{G_F}{\sqrt2}
 \sum_{q_1,q_2=d,s}V_{cq_1}V_{uq_2}^{*}
 \left(C_1Q_{1,c}^{q_1q_2u}
      +C_2Q_{2,c}^{q_1q_2u}\right)
 +{\cal H}_{\rm SL}^{(c)}+{\rm h.c.},
 \label{eq:heff-c}
\end{align}
where the penguin contribution to charm decay is CKM suppressed.  With 
\(\Gamma_\mu=\gamma_\mu(1-\gamma_5)\),
the  current--current operators are
\begin{align}
 Q_{1,Q}^{q_1q_2q_3}
 &=(\bar Q_{\alpha }  \Gamma_\mu q_{1\alpha}  )
   (\bar q_{2\beta }  \Gamma^\mu q_{3\beta } ) ,&
 Q_{2,Q}^{q_1q_2q_3}
 &=(\bar Q_{\beta }  \Gamma_\mu q_{1\alpha}  )
   (\bar q_{2\alpha }  \Gamma^\mu q_{3\beta } ).
 \label{eq:weak-current-operators}
\end{align}
Here \(\alpha\) and \(\beta\) are color indices\footnote{We  use  the opposite
\(Q_1,Q_2\) labels comparing to Ref.~\cite{Dulibic:2026}; all coefficient formulas  there should be
translated by \(C_1\leftrightarrow C_2\) before they are used below.}. 
We use the dominant
nonleptonic channels as representative examples; the penguin and semileptonic
Hamiltonians are given in Refs.~\cite{Dulibic:2026,Buchalla:1995vs}.
{\color{black}
In the numerical analysis, we fix the renormalization scales of the Wilson
coefficients at $\mu_b=4.4~\GeV$, $\mu_c=1.5~\GeV$, and
$\mu_{bc}=\sqrt{\mu_b\mu_c}$; the values of $C_{1,2}$ are given in
Table~\ref{tab:wilson-scale-inputs}. The hadronic-scale variation is implemented
through the matrix elements  in Sec.~\ref{sec:inputs}.}

\begin{table}[!th]
\centering
\color{black}
\caption{Values of $C_{1,2}$ at LO and NLO. The uncertainties arise from
$\alpha_s(M_Z)=0.1180\pm0.0009$~\cite{PDG}.}
\label{tab:wilson-scale-inputs}
\footnotesize
\vspace{0.2cm}
\setlength{\tabcolsep}{5pt}
\begin{tabular}{lccc}
\toprule
 Order & Scale & \(C_1\) & \(C_2\)  \\
\midrule
NLO & \(\mu_b\)    & \(1.078\pm0.001\) & \(-0.185\pm0.003\) \\
NLO & \(\mu_c\)    & \(1.184\pm0.005\) & \(-0.374\pm0.008\) \\
LO  & \(\mu_b\)    & \(1.101\pm0.001\) & \(-0.235\pm0.003\) \\
LO  & \(\mu_c\)    & \(1.189\pm0.003\) & \(-0.395\pm0.006\) \\
LO  & \(\mu_{bc}\) & \(1.138\pm0.002\) & \(-0.306\pm0.004\) \\
\bottomrule
\end{tabular}
\end{table}

The inclusive width 
of a hadron $H$ 
follows from the optical theorem
\begin{equation}
 \Gamma(H)=\frac{1}{2M_H}\langle H|{\cal T}|H\rangle,
 \qquad
 {\cal T}={\rm Im}\,i\!\int d^4x\,
 T\{ {\cal H}_{\rm eff}(x){\cal H}_{\rm eff}(0)\}.
 \label{eq:optical}
\end{equation}
At \({\cal O}(G_F^2)\),  there is no interference between
\({\cal H}_{\rm eff}^{(b)}\) and \({\cal H}_{\rm eff}^{(c)}\).  The transition operator has the schematic local expansion
\begin{align}
 {\cal T} 
 & =
 \sum _{Q= b,c } 
  \frac{G_F^2m_Q^5}{192\pi^3}  \left[
 \left({\cal C}_{3,Q}\bar QQ
      +\frac{{\cal C}_{5,Q}}{m_Q^2}{\cal O}_5^Q
      +\frac{c_{D,Q}}{m_Q^3}O_D^Q+\cdots\right)_{\!2q}
 \right.\nonumber\\[-1mm]
 &\hspace{2.3cm}\left.
 +16\pi^2\sum_{f=u,d,s}\left(
       \frac{\widetilde c_{6,i}^{\,Qf}}{m_Q^3}O_i^{Qf}
      +\frac{\widetilde c_{7,i}^{\,Qf}}{m_Q^4}R_i^{Qf}
      +\cdots\right)_{\!4q}\right] + {\cal T} _{bc} , 
 \label{eq:hqe-operator-expansion}
\end{align}
where \({\cal T}_{bc}\) contains transition operators involving both the
\(b\) and \(c\) quarks.
Repeated operator labels \(i\) are summed, and CKM and phase-space factors
are absorbed into the coefficients.  To display the leading two-quark
coefficients explicitly, consider a tree-level nonleptonic channel with
one massive quark in the final state.  The cases relevant here are
\(c\to s u\bar d\) and \(b\to c\bar u d\).  Define
\begin{equation}
 (Q,q,\xi_Q)=(c,s,|V_{cs}|^2|V_{ud}|^2)
 \quad\hbox{or}\quad
 (b,c,|V_{cb}|^2|V_{ud}|^2),
 \qquad x_q=\frac{m_q^2}{m_Q^2}.
\end{equation}
In the operator convention
\({\cal O}_5^Q=\bar Q\,\sigma_{\mu\nu}g_sG^{\mu\nu}Q/2\), the LO
coefficients are
\begin{align}
 {\cal C}_{3,Q}^{\rm non}
 &=
 \xi_Q\left(N_cC_1^2+N_cC_2^2+2C_1C_2\right)
 \left[(1-x_q^2)(1-8x_q+x_q^2)-12x_q^2\ln x_q\right],
 \label{eq:leading-c3}\\
 {\cal C}_{5,Q}^{\rm non} 
 &=
 -2 \xi_Q\left(N_cC_1^2+N_cC_2^2+2C_1C_2\right)(1-x_q)^4
 -16 \xi_Q C_1C_2(1-x_q)^3 .
 \label{eq:leading-c5}
\end{align}
Equations~\eqref{eq:leading-c3} and \eqref{eq:leading-c5} give the
one-mass nonleptonic contribution; the total \({\cal C}_{3,Q}\) and
\({\cal C}_{5,Q}\) are obtained by adding the remaining nonleptonic and
semileptonic channels with their corresponding CKM and mass factors
\cite{Bagan:1994zd,Bagan:1994qw,Krinner:2013cja,Mannel:2015jka,
Mannel:2023zei}.  The kinetic correction enters through the heavy-quark
bilinear,
\begin{equation}
\langle \overline Q Q \rangle _{{\cal B}_{bc}} 
 =1-\frac{\mu_{\pi,Q}^2-\mu_{G,Q}^2}{2m_Q^2}
 +O(1/m_Q^3),\qquad Q=b,c,
 \label{eq:bilinear-matrix-element}
\end{equation}
where
$
\langle   O  \rangle _{{\cal B}_{bc}}  \equiv
 {\langle {\cal B}_{bc}| O |{\cal B}_{bc}\rangle}/{2M_{{\cal B}_{bc}}}$. The bilinear
sector therefore contains the partonic, kinetic $\mu_{\pi ,Q}^2$, chromomagnetic $\mu_{G ,Q}^2$, and Darwin
contributions~$O_D^Q$~\cite{Bigi:1992su,Bigi:1993ex,Lenz:2020oce,King:2021xqp}.

\begin{figure}[t]
	\centering
	\begin{minipage}{0.225\linewidth}
		\centering
		\includegraphics[width=\linewidth]{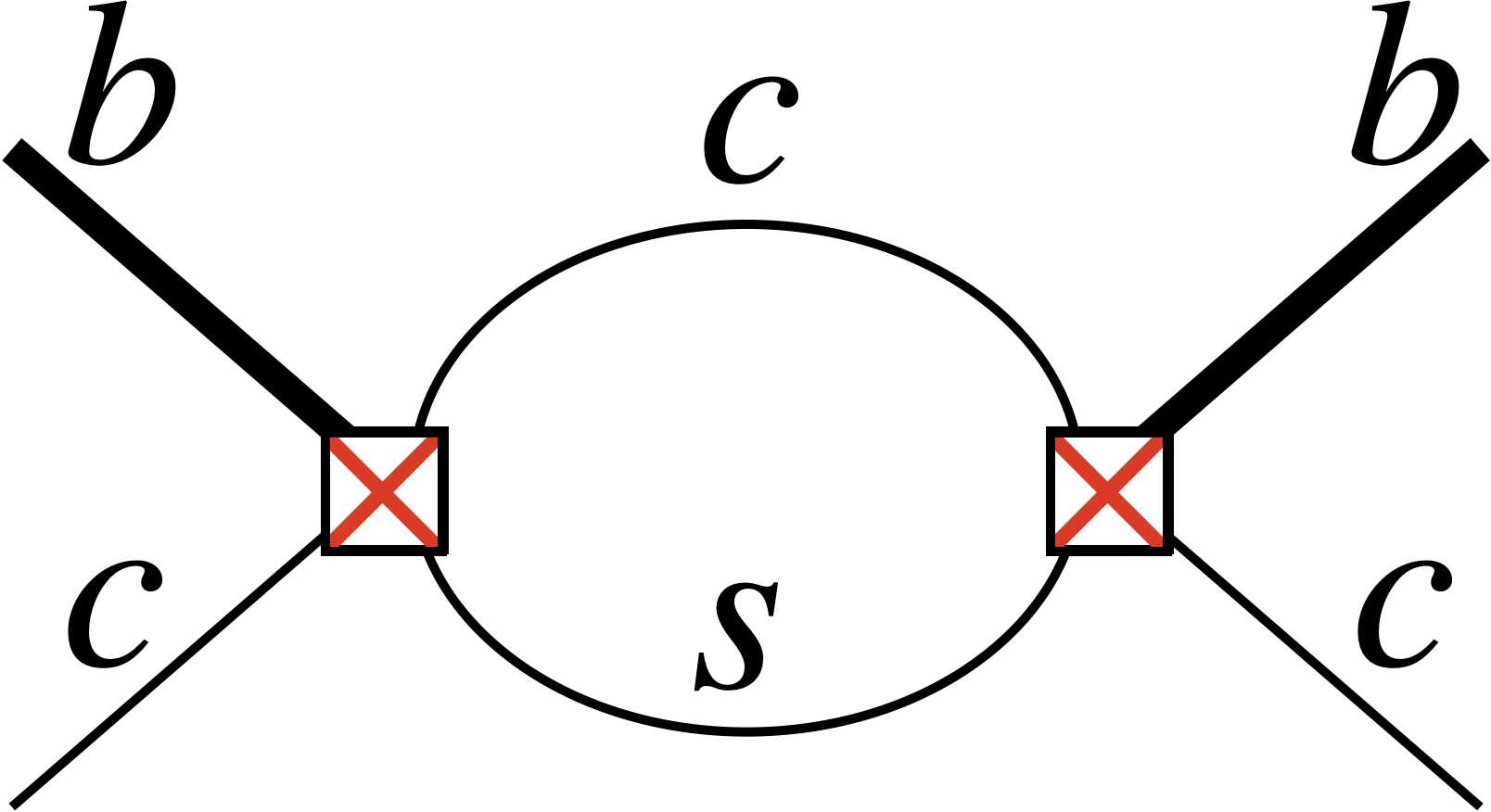}\\[-0.3em]
		(a)
	\end{minipage}\hfill
	\begin{minipage}{0.225\linewidth}
		\centering
		\includegraphics[width=\linewidth]{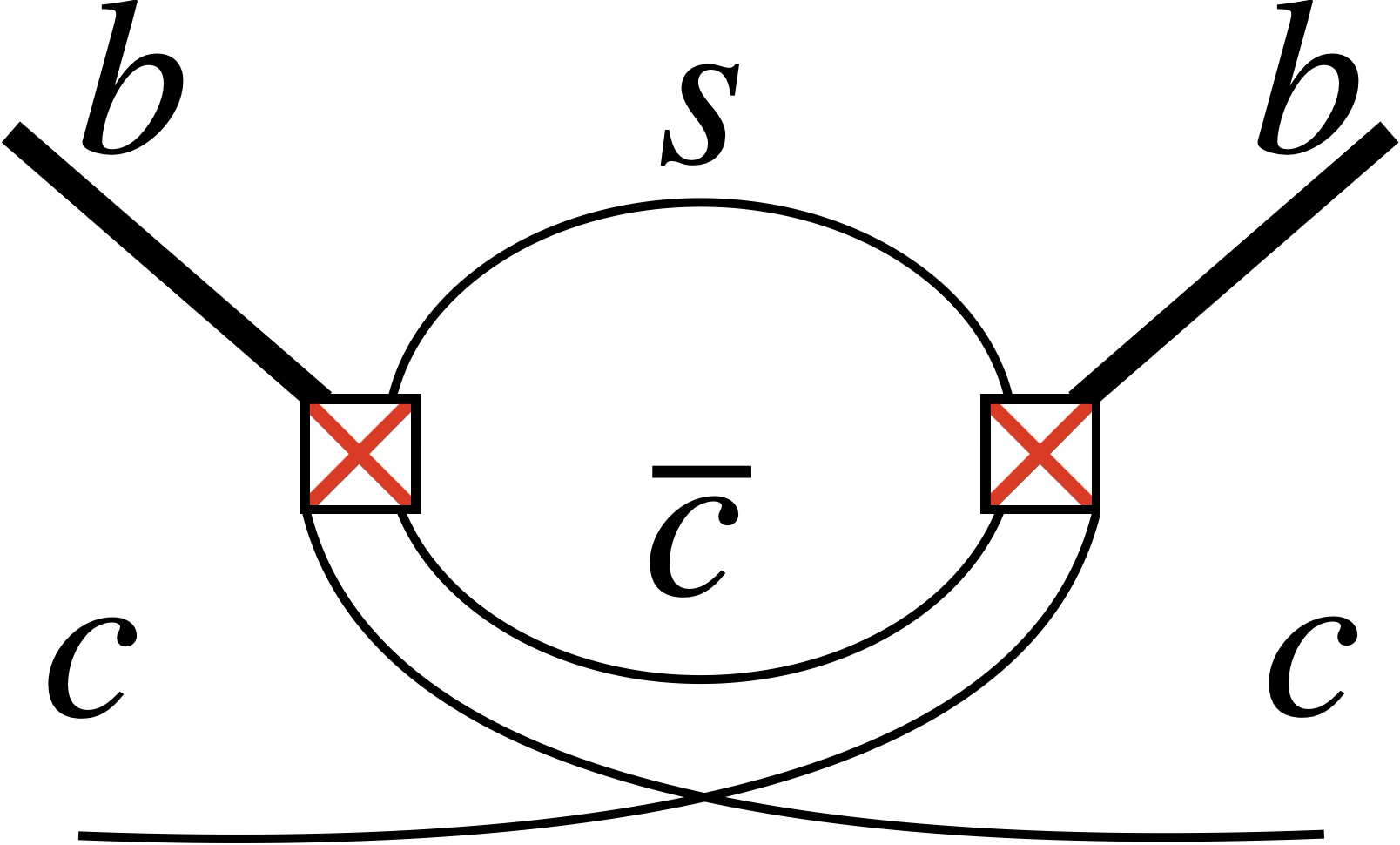}\\[-0.3em]
		(b)
	\end{minipage}\hfill
	\begin{minipage}{0.225\linewidth}
		\centering
		\includegraphics[width=\linewidth]{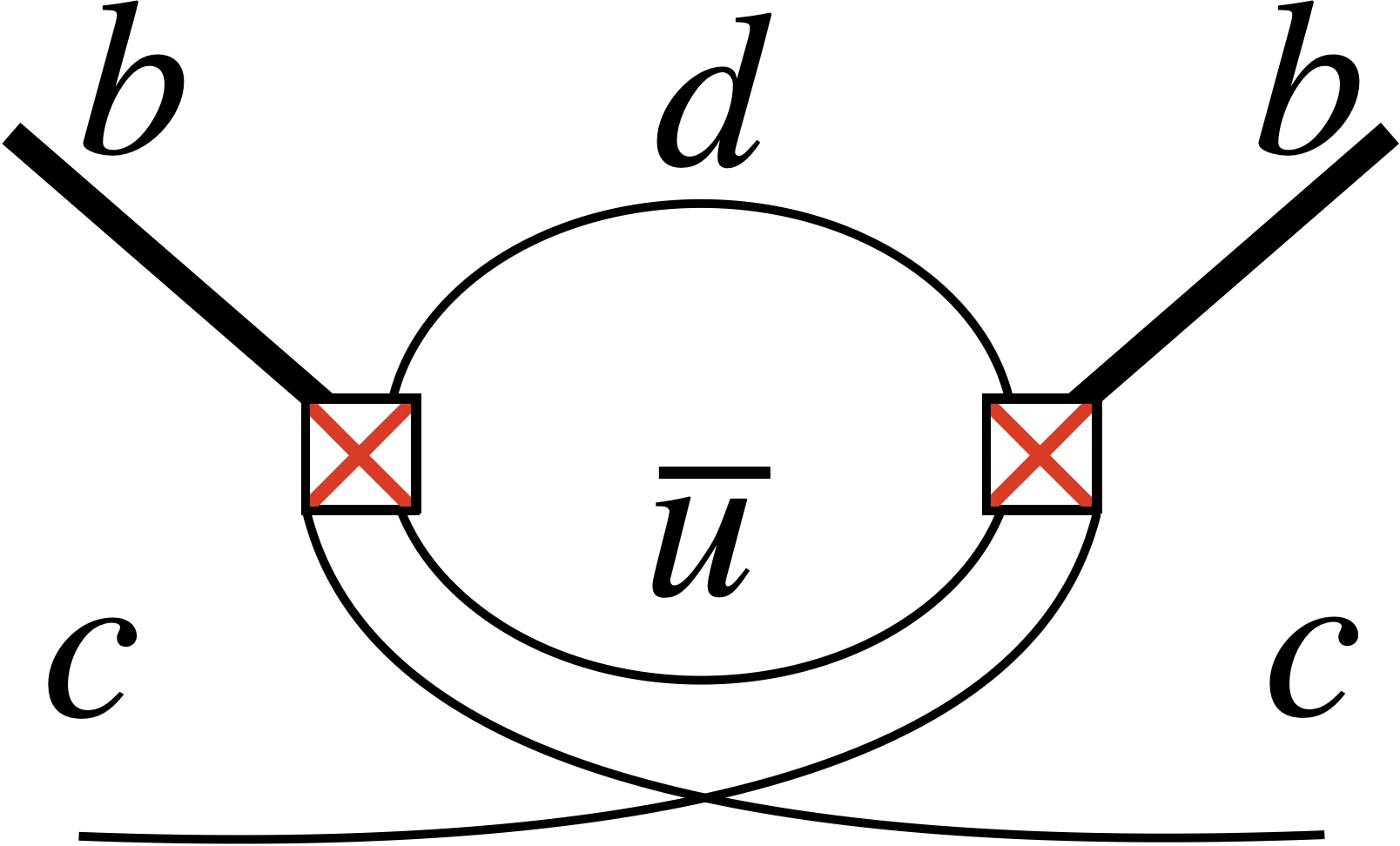}\\[-0.3em]
		(c)
	\end{minipage}\hfill
	\begin{minipage}{0.225\linewidth}
		\centering
		\includegraphics[width=\linewidth]{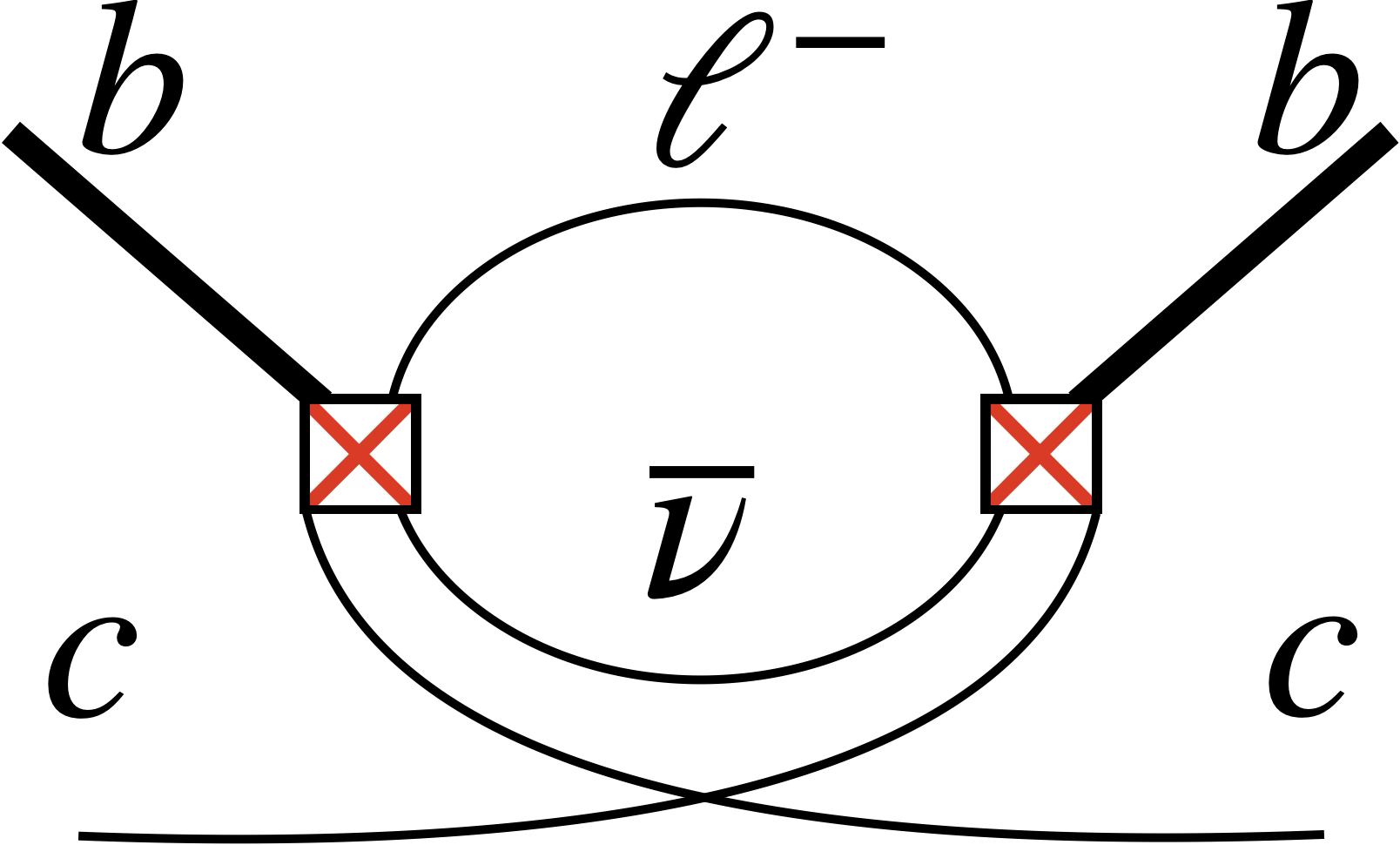}\\[-0.3em]
		(d)
	\end{minipage}
	\caption{Forward-scattering diagrams entering Eq.~\eqref{eq:tbc-tree}:
		(a) $W$-exchange; (b) finite-charm hadronic Pauli interference;
		(c) massless-\(u\) hadronic Pauli interference; and (d) semileptonic
		Pauli interference.  The red crosses denote insertions of the effective
		weak Hamiltonian.}
	\label{fig:tbc-spectator}
\end{figure}

The dimension-six spectator operators are suppressed by \(1/m_Q^3\), but their
two-body cut supplies the \(16\pi^2\) enhancement that drives lifetime
splittings \cite{Lenz:2014jha,Albrecht:2024oyn}.
The   heavy-light operators 
$O^{Qf}$ 
with \(Q=b,c\) and \(f=u,d,s\)
are the same as in our previous analysis~\cite{Cheng:2026mlv}.  The new active pair is \(Q=b\), \(f=c\).  For this pair the
left-handed vector structures are
\begin{align}
 O_1^{bc}
 &=(\bar b_\alpha\Gamma_\mu c_\alpha)
   (\bar c_\beta\Gamma^\mu b_\beta),~~~
 \widetilde O_1^{bc}
  =(\bar b_\alpha\Gamma_\mu c_\beta)
   (\bar c_\beta\Gamma^\mu b_\alpha).
 \label{eq:operators}
\end{align} 
The scalar structures for the same \(bc\) pair are
\begin{align}
 S_{ \pm \pm }^{bc}
 &=(\bar b^{\,\alpha}(1 \pm \gamma_5)c^{\,\alpha})
   (\bar c^{\,\beta}(1  \pm \gamma_5)b^{\,\beta}),\nonumber\\
 \widetilde S_{\pm \mp }^{bc}
 &=(\bar b^{\,\alpha}(1 \pm \gamma_5)c^{\,\beta})
   (\bar c^{\,\beta}(1 \mp \gamma_5)b^{\,\alpha}).
 \label{eq:bc-scalar-operators}
\end{align}  
At the leading order in heavy quark expansion
we have that 
\begin{equation}\label{211}
\langle S_{++}^{bc}\rangle
_{{\cal B}_{bc}}
=\langle S_{--}^{bc}\rangle_{{\cal B}_{bc}}
=\langle S_{+-}^{bc}\rangle_{{\cal B}_{bc}}=\langle S_{-+}^{bc}\rangle_{{\cal B}_{bc}} , 
\end{equation}
  with the
same relation for the color-rearranged operators.
It is due to that the pseudoscalar operators can be dropped at this order. 

The corresponding  transition operators are 
\begingroup
\allowdisplaybreaks[4]
\begin{subequations}
\label{eq:tbc-tree}
\begin{align}
 {\cal T}_{bc}^{\rm tree}
={}&{\cal T}_{bc}^{\rm WE}
+{\cal T}_{bc}^{\rm PI,c}
+{\cal T}_{bc}^{\rm PI,u}
+{\cal T}_{bc}^{\rm PI,SL},
\label{eq:tbc-tree-sum}\\
{\cal T}_{bc}^{\rm WE}
={}&|V_{cb}|^2\sum_{q=d,s}|V_{cq}|^2
\frac{G_F^2}{2\pi}(1-y_+^c)^2p_+^2
 \left[
 2C_1C_2O_1^{bc}
 +(C_1^2+C_2^2)\widetilde O_1^{bc}
 \right],
\label{eq:tbc-we}\\
{\cal T}_{bc}^{\rm PI,c}
={}&-|V_{cb}|^2\sum_{q=d,s}|V_{cq}|^2
\frac{G_F^2}{6\pi}(1-y_-^c)^2p_-^2
 \Bigg\{ 
 \left(N_cC_1^2+2C_1C_2\right)
 \Bigg\{
 \left(1+\frac{y_-^c}{2}\right)O_1^{bc}
 \nonumber\\[-1mm]
&\hspace{1.7cm}
-(1+2y_-^c)\frac{m_b^2}{p_-^2}
 \left[
 S_{-+}^{bc}-\frac{m_c}{m_b}
 \left(S_{--}^{bc}+S_{++}^{bc}\right)
 +\frac{m_c^2}{m_b^2}S_{+-}^{bc}
 \right]
 \Bigg\}
 \label{eq:tbc-pi-charm} 
  \\
& 
+C_2^2
 \Bigg\{
 \left(1+\frac{y_-^c}{2}\right)\widetilde O_1^{bc} 
-(1+2y_-^c)\frac{m_b^2}{p_-^2}
 \left[
 \widetilde S_{-+}^{bc}-\frac{m_c}{m_b}
 \left(\widetilde S_{--}^{bc}+\widetilde S_{++}^{bc}\right)
 +\frac{m_c^2}{m_b^2}\widetilde S_{+-}^{bc}
 \right]
 \Bigg\}
 \Bigg\}, \nonumber 
\\
{\cal T}_{bc}^{\rm PI,u}
={}&-|V_{cb}|^2\sum_{q=d,s}|V_{uq}|^2
 \frac{G_F^2}{6\pi}p_-^2
 \Bigg\{ 
 C_2^2
 \Bigg\{
 \widetilde O_1^{bc} 
 -\frac{m_b^2}{p_-^2}
 \left[
 \widetilde S_{-+}^{bc}-\frac{m_c}{m_b}
 \left(\widetilde S_{--}^{bc}+\widetilde S_{++}^{bc}\right)
 +\frac{m_c^2}{m_b^2}\widetilde S_{+-}^{bc}
 \right]
 \Bigg\} 
 \nonumber\\[-1mm]
&\!\!\!\!\! \!\!\! \!\!\!\!\!+
 \left(N_cC_1^2+2C_1C_2\right)
 \Bigg\{
 O_1^{bc} 
 -\frac{m_b^2}{p_-^2}
 \left[
 S_{-+}^{bc}-\frac{m_c}{m_b}
 \left(S_{--}^{bc}+S_{++}^{bc}\right)
 +\frac{m_c^2}{m_b^2}S_{+-}^{bc}
 \right]
 \Bigg\} \Bigg\} , 
\label{eq:tbc-pi-up}\\
{\cal T}_{bc}^{\rm PI,SL}
={}&-|V_{cb}|^2\sum_{\ell=e,\mu,\tau}
 \frac{G_F^2}{6\pi}(1-y_-^\ell)^2p_-^2
 \Bigg\{
 \left(1+\frac{y_-^\ell}{2}\right)O_1^{bc}
 \nonumber\\[-1mm]
&\hspace{0.7cm}
-(1+2y_-^\ell)\frac{m_b^2}{p_-^2}
 \left[
 S_{-+}^{bc}-\frac{m_c}{m_b}
 \left(S_{--}^{bc}+S_{++}^{bc}\right)
 +\frac{m_c^2}{m_b^2}S_{+-}^{bc}
 \right]
 \Bigg\} .
\label{eq:tbc-pi-sl}
\end{align}
\end{subequations}
\endgroup
Here \(p_\pm^2=(m_b\pm m_c)^2\) and
\(y_\pm^i=m_i^2/p_\pm^2\).   Equations~\eqref{eq:tbc-we}, \eqref{eq:tbc-pi-charm},
\eqref{eq:tbc-pi-up}, and \eqref{eq:tbc-pi-sl} correspond, respectively, to
panels (a)--(d) of Fig.~\ref{fig:tbc-spectator}.     Equation~\eqref{eq:tbc-tree} displays the complete
tree-level current--current contribution; penguin terms are included in
the numerical kernel but are not explicitly displayed here
\cite{Chang:2000ac,Cheng:2019sxr,Aebischer:2021ilm,Dulibic:2026}.
The minus signs multiplying \(S_{--}^{bc}\) and \(S_{++}^{bc}\), and their
color-rearranged analogues, in Eqs.~\eqref{eq:tbc-pi-charm} and
\eqref{eq:tbc-pi-up} follow from the on-shell equations of motion\footnote{ The corresponding two signs printed as positive in Eq.~(A.5) of
Ref.~\cite{Dulibic:2026} are therefore typos. }.  
 With the heavy-quark
relation  
in Eq.~\eqref{211}, the formulae here  reduce to the one  in Ref.~\cite{Cheng:2019sxr}.

The dimension-seven heavy-light contributions for \(Q=b,c\) and
\(f=u,d,s\) use the same four-operator definitions and ordering as in our
previous analysis~\cite{Cheng:2026mlv}.  We denote them by
\(R_i^{Qf}\):
\begin{align}
 R_1^{Qf}
 &=m_f(\bar Q_\alpha(1-\gamma_5)f_\alpha)
          (\bar f_\beta(1-\gamma_5)Q_\beta),\nonumber\\
 R_2^{Qf}
 &=m_f(\bar Q_\alpha(1+\gamma_5)f_\alpha)
          (\bar f_\beta(1+\gamma_5)Q_\beta),\nonumber\\
 R_3^{Qf}
 &=\frac{1}{m_Q}
   (\bar Q_\alpha\overleftarrow D_\rho\gamma_\mu(1-\gamma_5)D^\rho f_\alpha)
   (\bar f_\beta\gamma^\mu(1-\gamma_5)Q_\beta),\nonumber\\
 R_4^{Qf}
 &=\frac{1}{m_Q}
   (\bar Q_\alpha\overleftarrow D_\rho(1-\gamma_5)D^\rho f_\alpha)
   (\bar f_\beta(1+\gamma_5)Q_\beta),
 \label{eq:dimension-seven-basis}
\end{align}
together with the color-rearranged
\(\widetilde R_{1,\ldots,4}^{Qf}\)
\cite{Lenz:2013aua,Gabbiani:2003pq,Cheng:2018mwu,
Dulibic:2023jeu}.  These operators are denoted by \(P_{1,\ldots,4}\) in our
previous analysis.   
We retain these terms for both bottom--light and charm--light spectator
interactions.  Dimension-seven corrections to the heavy--heavy
\({\cal T}_{bc}\) term are not included.

\begin{table}[ht]
\centering
\caption{Status of the short-distance ingredients represented
in the common effective kernel.   }
\label{tab:coefficient-status}  
\footnotesize
\vspace{0.2cm}
\begin{tabular}{lll}
\toprule
HQE term &  Status & References \\
\midrule
Partonic \({\cal C}_{3,Q}\) &  NNLO
& \cite{Bagan:1994zd,Bagan:1994qw,Krinner:2013cja,
Fael:2024q2,Egner:2024nnlo,Dulibic:2026}\\
Chromomagnetic \({\cal C}_{5,Q}\) & NLO
& \cite{Mannel:2015jka,Mannel:2023zei,Mannel:2024bcud,Mannel:2025bccs}\\
Darwin & LO  
& \cite{Lenz:2020oce,King:2021xqp,Moreno:2022goo,Moreno:2024darwin}\\
Semileptonic power terms &   NLO 
& \cite{Falk:1994,Mannel:2017,Moreno:2022goo,Moreno:2024darwin}\\
Heavy-light dimension six & NLO
& \cite{Ciuchini:2001vx,Franco:2002fc,Beneke:2002rj}\\
Heavy-heavy \(bc\), dimension six & LO 
& \cite{Chang:2000ac,Cheng:2019sxr,Aebischer:2021ilm,Dulibic:2026}\\
Heavy-light dimension seven & LO
& \cite{Lenz:2013aua,Gabbiani:2003pq}\\
\bottomrule
\end{tabular}
\end{table}

In the numerical analysis, the partonic coefficients
\({\cal C}_{3,Q}\) are included through NNLO.  The chromomagnetic coefficient,
the semileptonic power terms, and the dimension-six heavy--light coefficients
are included at NLO, whereas the Darwin term, the heavy--heavy
\({\cal T}^{bc}\) contribution, and the dimension-seven coefficients are kept
at LO.  The corresponding short-distance kernels and references are
summarized in Table~\ref{tab:coefficient-status}.
Eqs.~\eqref{eq:leading-c3}, \eqref{eq:leading-c5}, and
\eqref{eq:tbc-tree} display the relevant LO structures for transparency; the
quoted perturbative corrections enter through the numerical short-distance
kernels.
Because \(m_c^2|V_{cs}|^2\gg m_b^2|V_{cb}|^2\), the leading spectator
effects are charm induced and follow the pattern
\(\Gamma_{\rm WE}^{cd}>\Gamma_{{\rm PI}^{+}}^{cs}>
|\Gamma_{{\rm PI}^{-}}^{cu}|\): \(cd\) $W$-exchange in \(\Xibcz\), \(cs\)
constructive Pauli interference in \(\Omegabcz\), and \(cu\) destructive
Pauli interference in \(\Xibcp\), respectively.  The same \(q=d\)
spectator-operator channel appears as destructive Pauli interference in
\(D^+\), where the truncated HQE width can become negative, but as
constructive $W$-exchange in the baryon~\cite{Cheng:2019sxr,Cheng:2026mlv}.

\section{Hadronic inputs for the mixing-angle analysis}
\label{sec:inputs}

We only need to calculate the parity-conserving~(PC) parts of the four-quark
operators for the spin-averaged forward matrix elements of positive-parity
baryons at rest; the vector--axial and scalar--pseudoscalar cross terms are
parity odd and vanish.  For an ordered active pair \(Qf\), the required
parity-conserving structures can be decomposed as
\begin{equation}
\begin{aligned}
 L^{Qf}={}&
 (\bar Q_\alpha\gamma_\mu f_\alpha)
 (\bar f_\beta\gamma^\mu Q_\beta)
 {}+
 (\bar Q_\alpha\gamma_\mu\gamma_5 f_\alpha)
 (\bar f_\beta\gamma^\mu\gamma_5 Q_\beta),\\
 S^{Qf}= {}&
 (\bar Q_\alpha f_\alpha)(\bar f_\beta Q_\beta),\qquad
 P^{Qf}=
 (\bar Q_\alpha\gamma_5 f_\alpha)
 (\bar f_\beta\gamma_5 Q_\beta),\\
 \widetilde L^{Qf}= {}&
 (\bar Q_\alpha\gamma_\mu f_\beta)
 (\bar f_\beta\gamma^\mu Q_\alpha)
 {}+
 (\bar Q_\alpha\gamma_\mu\gamma_5 f_\beta)
 (\bar f_\beta\gamma^\mu\gamma_5 Q_\alpha),\\
 \widetilde S^{Qf}= {}&
 (\bar Q_\alpha f_\beta)(\bar f_\beta Q_\alpha),\qquad
 \widetilde P^{Qf}=
 (\bar Q_\alpha\gamma_5 f_\beta)
 (\bar f_\beta\gamma_5 Q_\alpha).
\end{aligned}
\label{eq:parity-conserving-operators}
\end{equation}
The PC parts of the transition operators give
\begin{equation}
 [O_1^{Qf}]_{\rm PC}=L^{Qf},\quad
 [S_{\mp \pm }^{Qf}]_{\rm PC}=S^{Qf}-P^{Qf},
 \quad
 [S_{\pm \pm }^{Qf}]_{\rm PC}= S^{Qf}+P^{Qf} ,
\end{equation} 
with identical identities for the tilded operators.

The spin coupling coefficients can be evaluated independently of the spatial
wave functions in either HBM or NRQM. We first couple the spins of the \(b\)
and \(c\) quarks to \(S_{bc}=0\) or \(1\), and then couple the resulting
diquark spin to the light-quark spin to obtain total \(J=1/2\). With the quark
order \((b,c,q)\) in \({\cal B}_{bc}\), the corresponding spin wave functions
\(|S,m\rangle\) are given by
\begin{align}
 |1, \tfrac12\rangle
 &=\sqrt{\frac23}\,|\uparrow_b\uparrow_c\downarrow_q\rangle
 -\frac{
 |\uparrow_b\downarrow_c\uparrow_q\rangle+
 |\downarrow_b\uparrow_c\uparrow_q\rangle}{\sqrt6},
 ~~~~ |0, \tfrac12\rangle
  = \frac{
 |\uparrow_b\downarrow_c\uparrow_q\rangle-
 |\downarrow_b\uparrow_c\uparrow_q\rangle}{\sqrt2}, 
 \label{eq:spin-phase-convention}
\end{align}
where
$S$ is the spin of  the $bc$ diquark  and 
 $m$ is the baryon spin projection along $\hat z$.
These states can be written compactly as 
\begin{equation}
 |S,m\rangle
=\sum_{\boldsymbol\lambda} 
{\cal A}_{\boldsymbol\lambda} ^{Sm}
|\lambda_b\lambda_c\lambda_q\rangle, 
\end{equation} 
where \({\cal A}^{Sm}_{\boldsymbol\lambda}\) is the Clebsch--Gordan
amplitude in the fixed baryon order
\(\boldsymbol\lambda=(\lambda_b,\lambda_c,\lambda_q)\).  
For instance, we have that 
\begin{equation}
{\cal A}_ {\uparrow \downarrow\uparrow }
^{0 \frac12}
= 
 - {\cal A}_ {\downarrow \uparrow \uparrow }
^{0 \frac12}  = \frac{1}{\sqrt{2}},
\end{equation}
while other elements vanish. 

Throughout this discussion, bold symbols denote the \(2\times2\) matrices in
the ordered basis \((|1\rangle,|0\rangle)\); in particular,
\(\boldsymbol{\mu}_{\pi,Q}^2\equiv
\mu_{\pi,Q}^2\boldsymbol{1}_2\).  Since the two basis states differ only in
their spin coupling, the spin-independent kinetic correction
\(\mu_{\pi,Q}^2\) is common to \(|1\rangle\) and \(|0\rangle\).
In terms of the normalization
\begin{equation}
 C_{{\cal B}_{bc}Q}(\mu_H)\equiv
 \langle1|\boldsymbol{\mu}_{G,Q}^{2}(\mu_H)|1\rangle .
\end{equation}
the chromomagnetic matrices read
\begin{align}
 \boldsymbol{\mu}_{G,b}^{2}(\mu_H)
 &=C_{{\cal B}_{bc}b}(\mu_H)
 \begin{pmatrix}
  1&\sqrt3/2\\
  \sqrt3/2&0
 \end{pmatrix},
 \nonumber\\[2pt]
 \boldsymbol{\mu}_{G,c}^{2}(\mu_H)
 &=C_{{\cal B}_{bc}c}(\mu_H)
 \begin{pmatrix}
  1&-\sqrt3/2\\
  -\sqrt3/2&0
 \end{pmatrix}.
\label{eq:chromomagnetic-invariant-amplitude}
\end{align}
These matrices contain only the heavy--light Pauli term; the heavy--heavy
chromomagnetic interaction between the \(b\) and \(c\) quarks, proportional
to \(1/(m_bm_c)\), is omitted.

A normalized
four-quark matrix element can be decomposed into spin-flavor overlap
coefficients and reduced spatial matrix elements as
\begin{equation}
\langle I^{Qf}\rangle_{ {\cal B}_{bc} } 
= {\cal C}^{Qf,{\cal B}_{bc} }_{\rm flip}\langle\bar I^{Qf}\rangle_{\rm flip}
+ {\cal C}^{Qf,{\cal B}_{bc}  }_{\rm unflip}\langle\bar I^{Qf}\rangle_{\rm unflip},
\qquad I=L,S,P ,
\label{eq:reduced-spin-decomposition}
\end{equation}
where
\(\langle I^{Qf}\rangle_{{\cal B}_{bc}} \equiv\langle {\cal B}_{bc} |I^{Qf}|{\cal B}_{bc} \rangle/(2M_{{\cal B}_{bc}} )\).
The barred quantities are the reduced spatial matrix elements after the
spin-flavor coefficients have been factored out.
For a given \(I^{Qf}\), with \(I=L,S,P\), denote the ket spins of the active quarks by
\((\lambda_Q,\lambda_f)\) and their bra spins by
\((\bar\lambda_Q,\bar\lambda_f)\). 
The
spin-flavor overlap
coefficients are
\begin{align}
 {\cal C}_{\rm unflip}^{Qf;s_1s_2}
 &=\frac12 \sum_{\bar\lambda_Q,\bar\lambda_f,\lambda_f,\lambda_Q}
 \sum_{m=\pm1/2}\sum_{\lambda_h}
\left[
{\cal A}_{\bar{\boldsymbol\lambda}}^{s_1m}
\right]^*
{\cal A}_{\boldsymbol\lambda}^{s_2m}  
 (\chi_{\bar\lambda_Q}^\dagger\chi_{\lambda_f})
 (\chi_{\bar\lambda_f}^\dagger\chi_{\lambda_Q}),
 \nonumber\\
 {\cal C}_{\rm flip}^{Qf;s_1s_2}\delta_{rs}
 &=\frac12 \sum_{\bar\lambda_Q,\bar\lambda_f,\lambda_f,\lambda_Q}
\sum_{m=\pm1/2}\sum_{\lambda_h}
\left[
{\cal A}_{\bar{\boldsymbol\lambda}}^{s_1m}
\right]^*
{\cal A}_{\boldsymbol\lambda}^{s_2m} 
 (\chi_{\bar\lambda_Q}^\dagger\sigma_r\chi_{\lambda_f})
 (\chi_{\bar\lambda_f}^\dagger\sigma_s\chi_{\lambda_Q}),
 \label{eq:spin-coefficient-definitions}
\end{align}
where \(\lambda_h\) denotes the spectator-quark spin, for which
\(\lambda_h=\bar\lambda_h\) is imposed;
\(\chi_\uparrow=(1,0)^T\) and \(\chi_\downarrow=(0,1)^T\).
Here ${\cal C}
_{\rm (un)flip} $ stands for the spin-flavor overlap with 
quarks being (un)flipped in the quark model.   
 For instance, we have that
 \begin{equation}
  {\cal C}_{\rm unflip}^{bc ;0 0 }
  = \frac{1}{2}\left(
\left[   {\cal A} _{ \uparrow\downarrow \uparrow} ^{0\frac12}  
\right] ^* 
{\cal A} _{
	\downarrow 
	 \uparrow \uparrow} ^{0\frac12}  
 + 
\left[    {\cal A} _{ \downarrow\uparrow  \uparrow} ^{0   \frac12}  
 \right] ^* 
 {\cal A} _{
 	\uparrow \downarrow 
 	 \uparrow} ^{0 \frac12}   
  + \left[   {\cal A} _{ \uparrow\downarrow   \downarrow } ^{0 - \frac12}  
  \right] ^* 
  {\cal A} _{
  	\downarrow 
  	\uparrow    \downarrow} ^{0 - \frac12}  
  + 
  \left[    {\cal A} _{ \downarrow\uparrow  \downarrow} ^{0 - \frac12}  
  \right] ^* 
  {\cal A} _{
  	\uparrow \downarrow 
  	\downarrow} ^{0 - \frac12}    
  \right) \,. 
 \end{equation}
  In the ordered basis \((|1\rangle,|0\rangle)\), the pair-dependent
coefficient matrices are
\begin{equation}
 \begin{aligned} 
 {\cal  C}_{\rm unflip}^{bq}
 &=\begin{pmatrix}
 -\frac12&-\frac{\sqrt3}{2}\\
 -\frac{\sqrt3}{2}&\frac12
 \end{pmatrix},\qquad
 {\cal  C}_{\rm unflip}^{cq}
 =\begin{pmatrix}
 -\frac12&\frac{\sqrt3}{2}\\
 \frac{\sqrt3}{2}&\frac12
 \end{pmatrix},\\[2pt]
 {\cal  C}_{\rm unflip}^{bc}
 &=\begin{pmatrix}1&0\\0&-1\end{pmatrix},\qquad
 {\cal  C} _{\rm flip}^{Qf}
 =\frac23
 \begin{pmatrix}1&0\\0& 1\end{pmatrix} 
 -\frac13{\cal C}_{\rm unflip}^{Qf}.
 \end{aligned}
 \label{eq:spin-coefficient-matrices}
\end{equation}
The last relation follows directly from the Pauli completeness identity
$
 \sum_{r=1}^{3}(\sigma_r)_{\alpha\beta}
 (\sigma_r)_{\gamma\delta}
 =2\delta_{\alpha\delta}\delta_{\gamma\beta}
 -\delta_{\alpha\beta}\delta_{\gamma\delta}. 
$
A physical state is a mixture of $|1\rangle$ and $|0\rangle$ and is
denoted by
$ |{\cal B}_{bc}(\phi_\B\rangle)
=(\cos\phi_{\cal B},\sin\phi_{\cal B}) $, see Eq.~(\ref{eq:Bbc}).
Hence,  the full matrix elements are
 \begin{equation}
 {\cal C}^{Qf,{\cal B}_{bc} }_{\rm (un)flip}  = 
( \cos  \phi_{\cal B} ,  
\sin  \phi _{\cal B}  ) 
{\cal C} ^{Qf} _ {\rm (un)flip}  
\left( \begin{array}{c} 
	\cos  \phi _{\cal B} \\ 
	\sin  \phi  _{\cal B}  
\end{array}
\right) 
.
 \end{equation}
 The results of the spin factors are listed in Table~\ref{tab:spin-angle-coefficients}.

\begin{table}[!ht]
\centering
\caption{Pair-dependent spin coefficients and their coherent-state
expectation values.  Here
\(c_{\cal B}=\cos(2\phi_{\cal B})\) and
\(s_{\cal B}=\sin(2\phi_{\cal B})\).}
\label{tab:spin-angle-coefficients}  
\small 
\renewcommand{\arraystretch}{1.12}
\vspace{0.7em}
\begin{tabular}{@{}llcc@{}}
\toprule
State &   \(Qf\) & \({\cal C}_{\rm unflip}^{Qf,{\cal B}_{bc} } \)
& \({\cal C}_{\rm flip}^{Qf,{\cal B}_{bc}} \)\\
\midrule
\(\Xi_{bc}\) & \(bq\)
& \(-\frac12c_{\cal B}-\frac{\sqrt3}{2}s_{\cal B}\)
& \(\frac23+\frac16c_{\cal B}+\frac{\sqrt3}{6}s_{\cal B}\)\\
\(\Xi_{bc}\) & \(cq\)
& \(-\frac12c_{\cal B}+\frac{\sqrt3}{2}s_{\cal B}\)
& \(\frac23+\frac16c_{\cal B}-\frac{\sqrt3}{6}s_{\cal B}\)\\
\(\Xi_{bc}\) & \(bc\)
& \(c_{\cal B}\) & \(\frac23-\frac13c_{\cal B}\)\\
\(\Omega_{bc}\) & \(bs\)
& \(-\frac12c_{\cal B}-\frac{\sqrt3}{2}s_{\cal B}\)
& \(\frac23+\frac16c_{\cal B}+\frac{\sqrt3}{6}s_{\cal B}\)\\
\(\Omega_{bc}\) & \(cs\)
& \(-\frac12c_{\cal B}+\frac{\sqrt3}{2}s_{\cal B}\)
& \(\frac23+\frac16c_{\cal B}-\frac{\sqrt3}{6}s_{\cal B}\)\\
\(\Omega_{bc}\) & \(bc\)
& \(c_{\cal B}\) & \(\frac23-\frac13c_{\cal B}\)\\
\bottomrule
\end{tabular}
\end{table}
 
 For the reduced spatial matrix elements, 
we first discuss the HBM, followed by the NRQM.
We recapitulate only the core equations here; further details of
the HBM construction and its applications can be found in
Ref.~\cite{Cheng:2023jpz}.
We start from a static spherical bag of radius \(R\).  The lowest mode of a
quark of flavor \(q\), mass \(m_q\), and spin projection \(\lambda\) is
\begin{equation}
 \varphi_{q\lambda}(\bm r)=
 \begin{pmatrix}
  u_q(r)\,\chi_\lambda\\
  i v_q(r)\,\bm\sigma\!\cdot\!\hat{\bm r}\,\chi_\lambda
 \end{pmatrix},
 \qquad
 \begin{aligned}
  u_q(r)&={\cal N}_q
  \sqrt{\frac{E_q+m_q}{E_q}}\,j_0(p_q r),\\
  v_q(r)&={\cal N}_q
  \sqrt{\frac{E_q-m_q}{E_q}}\,j_1(p_q r),
 \end{aligned}
 \label{eq:static-bag-mode}
\end{equation}
for \(r<R\), where \(E_q^2=p_q^2+m_q^2\).
Here $j_{0,1}$ are the spherical Bessel functions, and $N_q$ is the
normalization coefficient.
  The   boundary condition
fixes the   momentum through
\begin{equation}
 \tan(p_qR)=\frac{p_qR}{1-m_qR-E_qR}.
 \label{eq:static-bag-boundary}
\end{equation}

A product of three modes localized about a fixed bag center is not an
eigenstate of the total momentum and therefore contains spurious
center-of-mass motion.  For a baryon at rest, the HBM construction removes
this artifact by projecting the static state over all bag centers~\cite{DeGrand:1975cf,Cheng:2023jpz}:
\begin{equation}
 \Psi_{\rm HBM}^{s,m}(\bm r_b,\bm r_c,\bm r_h)
 ={\cal N}\int d^3x \,
 \Psi_{\rm static}^{s,m}
 (\bm r_b-\bm x,\bm r_c-\bm x,\bm r_h-\bm x).
 \label{eq:hbm-projected-state}
\end{equation} 
To perform the numerical evaluation we shall use
\begin{equation}
 R_{\Xi_{bc}}=4.09~\GeV^{-1},
 \qquad
 R_{\Omega_{bc}}=4.18~\GeV^{-1},
 \label{eq:hbm-radii}
\end{equation}
and
\begin{equation}
 (m_{u,d},m_s,m_c,m_b)=(0.001,0.279,1.641,5.093)~\GeV ,
 \label{eq:hbm-masses}
\end{equation}
from the spectrum framework in the bag model~\cite{Zhang:2021yul}.
{\color{black}
These radii correspond to the lower mixed states in Table~X of
Ref.~\cite{Zhang:2021yul}, with chromomagnetic mixing included.
We retain them for the central values and estimate the bag-radius
uncertainty by repeating the matrix-element calculation at
$R=4.60~\GeV^{-1}$ for both baryons.
This reference value is the $\Lambda_b$ bag radius~\cite{Zhang:2021yul,Cheng:2026mlv}.
We use the shifts obtained by changing the spectrum-derived radii to this
reference value to estimate how uncertainties in the bag radii of doubly
heavy baryons affect our predictions.}

The HBM and NRQM inputs at the hadronic
scale \(\mu_H\) are collected in Table~\ref{tab:two-quark-matrix-elements}.
For the HBM we calculate the two-quark matrix elements directly, following
the bag-model construction used for \(\Xi_{QQ}\)~\cite{Cheng:2026mlv}.
The single-heavy kinetic parameter is
\begin{equation}
 \mu_{\pi,Q}^2=p_Q^2, 
 \label{eq:hbm-kinetic-matrix-element}
\end{equation} 
while 
the chromomagnetic interactions give~\cite{Cheng:2026mlv}
\begin{equation}
 C_{{\cal B}_{bc}Q}^{\rm HBM}
 =16\pi\alpha_s m_Q\sum_a\int d^3x\,
 \vec B_{Q\uparrow}^{\,a}(\vec x)\!\cdot\!
 \vec B_{f\uparrow}^{\,a}(\vec x).
 \label{eq:hbm-chromomagnetic-field}
\end{equation}
Here \(a\) is a color index and \(\uparrow\) denotes spin up.
The color-magnetic fields \(\vec B_i^{\,a}\) are defined in
Ref.~\cite{DeGrand:1975cf}.  Since
\(\vec B_{Q\uparrow}^{\,a}=O(1/m_Q)\),
\(C_{{\cal B}_{bc}Q}^{\rm HBM}=O(m_Q^0)\) at the leading power, up to logarithms
and finite-mass overlap effects. 
The $\alpha_s$ in Eq.~\eqref{eq:hbm-chromomagnetic-field} is adopted from the mass spectra fit~\cite{Zhang:2021yul}. 
Together with
Eq.~\eqref{eq:chromomagnetic-invariant-amplitude}, this determines the HBM
heavy--light chromomagnetic matrix.

The heavy--light Darwin matrix is not an independent input.
The gluon equation of motion then gives
\begin{align}
\langle  O _{D}^Q
\rangle _{{\cal B}_{bc} }
 =4\pi\alpha_s(\mu_H)
&\sum _f  \bigg[
 -\frac18\langle L^{Qf}\rangle_{{\cal B}_{bc}}
 +\frac1{24}\langle\widetilde L^{Qf}\rangle_{{\cal B}_{bc}}\nonumber\\[-1mm]
 &+\frac14\langle S^{Qf}-P^{Qf}\rangle_{{\cal B}_{bc}}
 -\frac1{12}\langle\widetilde S^{Qf}
                    -\widetilde P^{Qf}\rangle_{{\cal B}_{bc}}
 \bigg]+O(1/m_Q).
\label{eq:darwin-four-quark-relation}
\end{align}
In the valence approximation used here, the flavor sum includes \(f=c\) when
\(Q=b\), \(f=u\) or \(d\) for the corresponding \(\Xi_{bc}\) charge state,
and \(f =s\) for \(\Omega_{bc}\).  With the
valence-color relation in
Eq.~\eqref{eq:color-rearranged-reduced-elements}, the HBM inputs give
heavy--light Darwin eigenvalues of \(0.014\)--\(0.023~\GeV^3\) at
\(\mu_H\).  Their extra contribution to the width is tiny.

In a forward matrix element, the bra and ket bag centers may be written as
\(\bm x\pm\bm x_\Delta/2\).  Integration over their average position
\(\bm x\) leaves the relative displacement \(\bm x_\Delta\); the spectator
\(h\) then contributes to the overlap \(D_h(\bm x_\Delta)\), while the
four-quark operator acts on the active pair \(Qf\). 
The remaining HBM calculation is needed here only for
\(\langle\bar I^{Qf}\rangle_{\rm flip}\) and
\(\langle\bar I^{Qf}\rangle_{\rm unflip}\).  All baryon-spin dependence has already been isolated in
the coefficients of Eq.~\eqref{eq:spin-coefficient-matrices}; the barred quantities
below contain only the displaced-bag spatial integrals.

For any constituent flavor \(q\), define the one-quark overlap
\begin{equation}
 \begin{aligned}
 D_q(\bm x_\Delta)
 &=\int d^3y\,
 \left[
 u_q(y^+)u_q(y^-)
 +v_q(y^+)v_q(y^-)
 \hat{\bm y}^{+}\!\cdot\!\hat{\bm y}^{-}
 \right],\\
 \bm y^\pm&=\bm y\pm\frac{\bm x_\Delta}{2},\qquad 
 \hat{\bm y}^\pm=\bm y^\pm/  y^\pm   ,\qquad y^\pm = | \bm y ^\pm | . 
 \end{aligned}
 \label{eq:hbm-spectator-overlap}
\end{equation}
The  normalization is fixed by
\begin{equation}
{\cal N}_{ {\cal B}_{bc} } 
 = 
 \int d^3x_\Delta\,
 D_b(\bm x_\Delta)D_c(\bm x_\Delta)D_h(\bm x_\Delta),
 \label{eq:hbm-state-normalization}
\end{equation}
which gives the normalized matrix elements of 
\begin{equation}
 {\cal J}_h[F]\equiv
\frac{1 }{{\cal N}_{ {\cal B}_{bc} } } 
 \int d^3x_\Delta\,D_h(\bm x_\Delta)
 \int d^3x\,F(\bm x_\Delta,\bm x),
 \label{eq:hbm-reduced-functional}
\end{equation}
where \(h\) is the spectator flavor.
For the active pair, let
\begin{equation}
 \begin{gathered}
 \bm x^\pm=\bm x\pm\frac{\bm x_\Delta}{2},\qquad
 r^\pm=|\bm x^\pm|,\qquad
 u_a^\pm=u_a(r^\pm),\quad v_a^\pm=v_a(r^\pm),
 \quad a\in\{Q,f\},\\
 c_x=\hat{\bm x}^{+}\!\cdot\!\hat{\bm x}^{-},\qquad
 w_x=\frac{|\,\bm x_\Delta\times\bm x\,|^2}{(r^+r^-)^2},
 \qquad
 V^{Qf}=v_Q^+v_f^-v_f^+v_Q^- ,
 \end{gathered}
 \label{eq:hbm-reduced-notation}
\end{equation}
and introduce
\begin{align}
 X_\pm^{Qf}
 &=u_Q^+u_f^-\mathbin{\pm}v_Q^+v_f^-c_x,&
 Y_\pm^{Qf}
 &=u_f^+u_Q^-\mathbin{\pm}v_f^+v_Q^-c_x,\nonumber\\
 \bm A_\pm^{Qf}
 &=u_Q^+v_f^-\hat{\bm x}^-
   \mathbin{\pm}v_Q^+u_f^-\hat{\bm x}^+,&
 \bm B_\pm^{Qf}
 &=u_f^+v_Q^-\hat{\bm x}^-
   \mathbin{\pm}v_f^+u_Q^-\hat{\bm x}^+ .
 \label{eq:hbm-reduced-building-blocks}
\end{align}
The final left-current term is abbreviated as
\begin{align}
 H^{Qf}\equiv{}&
 3u_Q^+u_f^-u_f^+u_Q^-
 +V^{Qf}(2+c_x^2) -\left(
 u_Q^+u_f^-v_f^+v_Q^-
 +v_Q^+v_f^-u_f^+u_Q^-
 \right)c_x .
 \label{eq:hbm-left-final-term}
\end{align}

The reduced scalar matrix elements are
\begin{align}
 \langle\bar S^{Qf}\rangle_{\rm flip}
 &=-{\cal J}_h[w_xV^{Qf}],&
 \langle\bar S^{Qf}\rangle_{\rm unflip}
 &={\cal J}_h[X_-^{Qf}Y_-^{Qf}].
 \label{eq:hbm-reduced-scalar}
\end{align}
For the pseudoscalar operator, we have
\begin{align}
 \langle\bar P^{Qf}\rangle_{\rm flip}
 &=-{\cal J}_h[\bm A_+^{Qf}\!\cdot\!\bm B_+^{Qf}],&
 \langle\bar P^{Qf}\rangle_{\rm unflip}
 &=0 .
 \label{eq:hbm-reduced-pseudoscalar}
\end{align}
Finally, the reduced left-handed-current matrix elements are
\begin{align}
 \langle\bar L^{Qf}\rangle_{\rm flip}
 &=-{\cal J}_h\!\left[
 w_xV^{Qf}
 +\bm A_-^{Qf}\!\cdot\!\bm B_-^{Qf}
 +2\bm A_+^{Qf}\!\cdot\!\bm B_+^{Qf}
 +H^{Qf}\right],\nonumber\\
 \langle\bar L^{Qf}\rangle_{\rm unflip}
 &={\cal J}_h\!\left[
 X_+^{Qf}Y_+^{Qf}
 +\bm A_-^{Qf}\!\cdot\!\bm B_-^{Qf}
 +w_xV^{Qf}\right].
 \label{eq:hbm-reduced-left}
\end{align} 
 Equations
\eqref{eq:hbm-reduced-scalar}--\eqref{eq:hbm-reduced-left}, combined with
the spin coefficients of Eq.~\eqref{eq:spin-coefficient-matrices}, therefore provide
all HBM four-quark matrix elements used below.

The Gaussian NRQM calculation is needed here only for
\(\langle\bar I^{Qf}\rangle_{\rm flip}\) and
\(\langle\bar I^{Qf}\rangle_{\rm unflip}\).   As in the HBM construction above, the
spin-flavor coefficients are kept separate from the reduced spatial
matrix elements. 
We use the Jacobi coordinates \cite{DeRujula:1975qlm}
\begin{equation}
 \bm\rho=\bm r_b-\bm r_c,\qquad
 \bm\lambda=
 \frac{m_b\bm r_b+m_c\bm r_c}{m_b+m_c}-\bm r_q ,
 \label{eq:jacobi}
\end{equation}
and the normalized Gaussian wave function
\begin{equation}
 \psi(\bm\rho,\bm\lambda)=
 \left(\frac{\alpha_\rho^2}{\pi}\right)^{3/4}
 \left(\frac{\alpha_\lambda^2}{\pi}\right)^{3/4}
 \exp\!\left[-\frac{\alpha_\rho^2\rho^2}{2}
             -\frac{\alpha_\lambda^2\lambda^2}{2}\right].
 \label{eq:nrqm-wavefunction}
\end{equation}
In the convention of Ref.~\cite{NRQM}, the \(bc\) oscillator parameters
are
\begin{align}
 \alpha_{\rho,bc}
 &=
 \alpha_{\rho,cc}
 \left[
 \frac{m_bm_c/(m_b+m_c)}{m_c/2}
 \right]^{1/4},\nonumber\\
 \alpha_{\lambda,bcq}
 &=
 \left[
 \frac{4m_q(m_b+m_c)^2}
 {3m_bm_c(m_b+m_c+m_q)}
 \right]^{1/4}\alpha_{\rho,bc}.
 \label{eq:oscillator-scaling}
\end{align}
The constituent quark masses and the single uncertainty-bearing oscillator input
are
\begin{align}
 &(m_{u,d},m_s,m_c,m_b)
 =(0.330,0.450,1.600,5.093)~\GeV ,
 \label{eq:nrqm-masses}\\
 &\alpha_{\rho,cc}=0.420\pm0.010~\GeV .
 \label{eq:nrqm-alpha-input}
\end{align}
The coherent propagation through Eq.~\eqref{eq:oscillator-scaling} gives
\begin{equation}
 \alpha_{\rho,bc}=0.466\pm0.011~\GeV,\qquad
 \alpha_{\lambda,bcq}=0.357\pm0.009~\GeV,\qquad
 \alpha_{\lambda,bcs}=0.385\pm0.009~\GeV .
 \label{eq:nrqm-derived-oscillators}
\end{equation}

As in Ref.~\cite{Cheng:2018mwu}, we identify
\(\mu_{\pi,Q}^2=\langle\bm p_Q^2\rangle\) and apply the Gaussian wave function to get
\begin{equation}
 \mu_{\pi,Q}^2
 =\frac32\left[
 \alpha_\rho^2+
 \left(\frac{m_Q}{m_b+m_c}\right)^2\alpha_\lambda^2
 \right],\qquad Q=b,c .
 \label{eq:nrqm-kinetic-matrix-elements}
\end{equation}
Using the contact-hyperfine relation of Ref.~\cite{Cheng:2018mwu}, we obtain
\begin{equation}
 C_{{\cal B}_{bc}Q}^{\rm NRQM}
 =\frac{16\pi\alpha_s}{9m_f}\,|\psi_{Qf}(0)|^2,
 \label{eq:nrqm-chromomagnetic-invariant}
\end{equation}
   For the Gaussian wave function, we have 
\begin{align}
 \!\!|\psi_{Qf}(0)|^2
\!\equiv\! 
\int  \! d^3\rho\,d^3\lambda\,
|\psi(\boldsymbol\rho,\boldsymbol\lambda)|^2
\delta^{(3)}\!\left( \mathbf r_Q -\mathbf r_f \right) 
 =\frac{1}{\pi^{3/2}}
 \left(
  \frac{1}{\alpha_\lambda^2}
  +\frac{m_b^2+m_c^2 - m_Q^2 }{(m_b+m_c)^2\alpha_\rho^2}
 \right)^{-\frac{3}{2}}.  
\end{align}
Here  \(\alpha_\rho=\alpha_{\rho,bc}\), and
\(\alpha_\lambda=\alpha_{\lambda,bcf}\). 
In the numerical analysis, we take the energy scale $\mu_H$ to lie in the range $0.8$--$1.0~\mathrm{GeV}$ in the $\overline{\mathrm{MS}}$ scheme, corresponding at five-loop order to $\alpha_s(0.8~\mathrm{GeV})=0.672$ and $\alpha_s(1.0~\mathrm{GeV})=0.417$, respectively~\cite{Herren:2017osy}.

\begin{table}[!ht]
\centering
\caption{HBM and NRQM invariant two-quark amplitudes at the hadronic scale
\(\mu_H\), in units of \(\GeV^2\).  The kinetic amplitudes are spin-independent, while \(C_{{\cal B}_{bc}Q}\), defined in
Eq.~\eqref{eq:chromomagnetic-invariant-amplitude}, determines the complete
heavy--light chromomagnetic matrix in the ordered basis
\((|1\rangle,|0\rangle)\).
{\color{black}The HBM uncertainties are the absolute shifts obtained with
$R=4.60~\GeV^{-1}$ relative to the spectrum-derived central radii,
symmetrized about the central values.}}
\label{tab:two-quark-matrix-elements}
\footnotesize
\vspace{6pt}
\setlength{\tabcolsep}{4pt}
\renewcommand{\arraystretch}{1.12}
\begin{tabular}{@{}llcc@{}}
\toprule
Model & Invariant amplitude & \(\Xi_{bc}\) & \(\Omega_{bc}\)\\
\midrule
\multirow{4}{*}{\(\HBM\)}
& \(\mu_{\pi,b}^2\) & {\color{black}\(0.562\pm0.115\)} & {\color{black}\(0.539\pm0.092\)}\\
& \(\mu_{\pi,c}^2\) & {\color{black}\(0.510\pm0.101\)} & {\color{black}\(0.490\pm0.080\)}\\
& \(C_{{\cal B}_{bc}b}\)
& {\color{black}\(0.150\pm0.020\)} & {\color{black}\(0.126\pm0.016\)}\\
& \(C_{{\cal B}_{bc}c}\)
& {\color{black}\(0.132\pm0.015\)} & {\color{black}\(0.110\pm0.012\)}\\
\midrule
\multirow{4}{*}{\(\NRQM\)}
& \(\mu_{\pi,b}^2\) & \(0.437\pm0.021\) & \(0.455\pm0.022\)\\
& \(\mu_{\pi,c}^2\) & \(0.337\pm0.016\) & \(0.339\pm0.016\)\\
& \(C_{{\cal B}_{bc}b}\)
& \(0.0654^{+0.0239}_{-0.0112}\)
& \(0.0593^{+0.0216}_{-0.0102}\)\\
& \(C_{{\cal B}_{bc}c}\)
& \(0.0443^{+0.0162}_{-0.0076}\)
& \(0.0381^{+0.0139}_{-0.0065}\)\\
\bottomrule
\end{tabular}
\end{table}

Because the NRQM spinors have no lower Dirac components, the reduced
left-handed-current matrix elements are
\begin{align}
 \langle\bar L^{Qf}\rangle_{\rm flip}
 &=-3Z_{Qf},&
 \langle\bar L^{Qf}\rangle_{\rm unflip}
 &=Z_{Qf}.
 \label{eq:nrqm-reduced-left}
\end{align}
The reduced scalar matrix elements are
\begin{align}
 \langle\bar S^{Qf}\rangle_{\rm flip}
 &=0,&
 \langle\bar S^{Qf}\rangle_{\rm unflip}
 &=Z_{Qf},
 \label{eq:nrqm-reduced-scalar}
\end{align}
whereas both reduced pseudoscalar matrix elements vanish:
\begin{align}
 \langle\bar P^{Qf}\rangle_{\rm flip}
 &=0,&
 \langle\bar P^{Qf}\rangle_{\rm unflip}
 &=0 .
 \label{eq:nrqm-reduced-pseudoscalar}
\end{align}
These six reduced quantities, combined with the spin coefficients of
Eq.~\eqref{eq:spin-coefficient-matrices}, give the NRQM four-quark matrix
elements used below.   

  Table~\ref{tab:matrix-elements}   lists
the six reduced spatial quantities
\(\langle\bar I^{Qf}\rangle_{\rm flip,unflip}\).
The complete spin-basis matrices, and hence the matrix element at any
mixing angle, are reconstructed from
Eq.~\eqref{eq:reduced-spin-decomposition} and the coefficient matrices in
Eq.~\eqref{eq:spin-coefficient-matrices}.  
The two models give similar values for $O^{bq}$.  Nevertheless, the NRQM
\(\langle\bar L^{cq}\rangle_{\rm flip}\) and
\(\langle\bar L^{cs}\rangle_{\rm flip}\) magnitudes are much  smaller, while its
heavy-heavy unflip overlap is 
significantly 
larger.    
In the valence three-quark approximation, the antisymmetric baryon color
wave function gives
\begin{equation}
 \langle\overline{\widetilde I}^{\,Qf}\rangle_{\rm (un)flip}  =-\langle\bar I^{Qf}\rangle_{\rm (un)flip}.
 \label{eq:color-rearranged-reduced-elements}
\end{equation}
Thus the color-rearranged matrix elements are fixed by an overall minus
sign and need not be listed independently.

\begin{table}[!ht]
\centering
\caption{Reduced flip and unflip matrix elements in the two spatial
models at the energy scale of $\mu_H$.  Entries are in \(10^{-3}\,\GeV^3\).  The color-rearranged
partners are obtained from
Eq.~\eqref{eq:color-rearranged-reduced-elements} and are not repeated.}
\label{tab:matrix-elements} 
\footnotesize
\setlength{\tabcolsep}{2.4pt}
\renewcommand{\arraystretch}{1.2}
\begin{tabular}{@{}lllcccccc@{}}
\toprule
State &  $Qf$  & model
& \multicolumn{2}{c}{\(\langle\bar L^{Qf}\rangle\)}
& \multicolumn{2}{c}{\(\langle\bar S^{Qf}\rangle\)}
& \multicolumn{2}{c}{\(\langle\bar P^{Qf}\rangle\)}\\
\cmidrule(lr){4-5}\cmidrule(lr){6-7}\cmidrule(lr){8-9}
& & & flip & unflip & flip & unflip & flip & unflip\\
\midrule
\multirow{6}{*}{\(\Xi_{bc}\)}
& \multirow{2}{*}{\(bq\)}
& \(\HBM\) & \(-22.7\) & \(7.2\) & \(0.0\) & \(6.8\) & \(-0.9\) & \(0.0\)\\
& & \(\NRQM\)
& \(-23.4\) & \(7.8\) & \(0.0\) & \(7.8\) & \(0.0\) & \(0.0\)\\
\cline{2-9}
& \multirow{2}{*}{\(cq\)}
& \(\HBM\) & \(-24.0\) & \(7.8\) & \(0.0\) & \(6.3\) & \(-1.7\) & \(0.0\)\\
& & \(\NRQM\)
& \(-15.9\) & \(5.3\) & \(0.0\) & \(5.3\) & \(0.0\) & \(0.0\)\\
\cline{2-9}
& \multirow{2}{*}{\(bc\)}
& \(\HBM\) & \(-34.3\) & \(11.4\) & \(0.0\) & \(11.2\) & \(-0.3\) & \(0.0\)\\
& & \(\NRQM\)
& \(-54.7\) & \(18.2\) & \(0.0\) & \(18.2\) & \(0.0\) & \(0.0\)\\
\midrule
\multirow{6}{*}{\(\Omega_{bc}\)}
& \multirow{2}{*}{\(bs\)}
& \(\HBM\) & \(-25.0\) & \(8.1\) & \(0.0\) & \(7.8\) & \(-0.7\) & \(0.0\)\\
& & \(\NRQM\)
& \(-28.9\) & \(9.6\) & \(0.0\) & \(9.6\) & \(0.0\) & \(0.0\)\\
\cline{2-9}
& \multirow{2}{*}{\(cs\)}
& \(\HBM\) & \(-25.4\) & \(8.4\) & \(0.0\) & \(7.1\) & \(-1.4\) & \(0.0\)\\
& & \(\NRQM\)
& \(-18.6\) & \(6.2\) & \(0.0\) & \(6.2\) & \(0.0\) & \(0.0\)\\
\cline{2-9}
& \multirow{2}{*}{\(bc\)}
& \(\HBM\) & \(-32.2\) & \(10.7\) & \(0.0\) & \(10.5\) & \(-0.3\) & \(0.0\)\\
& & \(\NRQM\)
& \(-54.7\) & \(18.2\) & \(0.0\) & \(18.2\) & \(0.0\) & \(0.0\)\\
\bottomrule
\end{tabular}
\end{table}

At the leading order in heavy quark expansion, only the time component of
the covariant derivative in the dimension-seven operators is retained.  In
the HBM, the light-quark wave function is an energy eigenstate, whereas in
the NRQM its energy is approximated by the constituent mass.  In the
four-operator convention of Eq.~\eqref{eq:dimension-seven-basis}, we use
\begin{equation}
 \langle R_3^{Qf}\rangle
 =E_f\langle L^{Qf}\rangle,
 \qquad
 \langle R_4^{Qf}\rangle
 =E_f\langle S^{Qf}-P^{Qf}\rangle,
\end{equation}
with \(E_{u,d}=0.33~\GeV\) and \(E_s=0.50~\GeV\), 
 and 
\begin{equation}
 \langle R_1^{Qs}\rangle
 =\langle R_2^{Qs}\rangle
 =m_s\langle S^{Qs}+P^{Qs}\rangle,
\end{equation}
where \(m_s=0.279~\GeV\) in the HBM and \(0.450~\GeV\) in the NRQM.
 
In the partonic calculation used to determine the HQE Wilson coefficients,
the initial heavy quark and the final-state quarks are taken to be on shell.  This
kinematic prescription does not, however, fix the heavy-quark mass scheme.
For the mass-sensitivity rescaling of the fixed kinetic-scheme kernels,
following our previous analysis for the numerical intervals and the standard
discussion of the pole-mass ambiguity~\cite{Cheng:2026mlv,Beneke:1998ui},
we use the intervals between the pole masses obtained from the one- and
two-loop conversions:
\begin{equation}
 m_c=1.59\pm0.09~\GeV,\qquad
 m_b=4.70\pm0.10~\GeV .
 \label{eq:mass-uncertainties}
\end{equation}
  The HBM and NRQM matrix elements at \(\mu_H\) instead use the
model-specific masses quoted in Eqs.~\eqref{eq:hbm-masses} and
\eqref{eq:nrqm-masses}.  For the charm mass entering the phase-space
dependence of \({\cal C}_{3,b}\), the imported kernel uses the running
\(\overline{\mathrm{MS}}\) input
\(\overline{m}_c(\mu _b )=0.91 ~\GeV\), following the standard inclusive
\(b\)-decay prescription~\cite{Krinner:2013cja,Dulibic:2026}.

 Finally, we collect the scale-evolution prescriptions for the two- and
four-quark matrix elements.
In this work, we keep the energy scale $\mu_Q$ of the HQE Wilson coefficients fixed while evolving the matrix elements from $\mu_H$ to $\mu_Q$. 
The kinetic parameter \(\mu_{\pi,Q}^2\) is protected by reparametrization
invariance and is therefore not evolved. For the chromomagnetic
matrix elements, the spectrum-matching prescription  is applied entry by entry~\cite{Cheng:2026mlv}:
\begin{align}
& C_G(m_Q,\mu_H)\boldsymbol{\mu}_{G,Q}^2(\mu_H)
 =C_G(m_Q,m_Q)\boldsymbol{\mu}_{G,Q}^2(m_Q),
 \nonumber\\
& C_G(m_Q,\mu_H)=1,\qquad
 C_G(m_b,m_b)=1.2664,\qquad C_G(m_c,m_c)=1.6506
 \label{eq:two-quark-chromomagnetic-evolution}
\end{align}
at $\mu_H=1$~GeV.
The latter two values are the three-loop hard-matching
coefficients~\cite{Grozin:2007fh}.
For the Wilson coefficients in the effective Hamiltonians of
eqs.~\eqref{eq:heff-b} and \eqref{eq:heff-c}, we evaluate
$C_1$ and $C_2$ at the scale $\mu_Q$ if only one of the heavy quark $Q$ is involved. For $O^{bc}$, we use the scale
$\mu_{bc}=\sqrt{\mu_b\mu_c}$. The scales used in our analysis are
\begin{equation}
 \mu_H=1.0\pm0.2~\GeV,
 \qquad \mu_c=1.5~\GeV,\qquad \mu_b=4.4~\GeV.
 \label{eq:scale-uncertainties}
\end{equation} 
For the dimension-seven operators and $O^{bc}$, for which NLO
corrections are not yet known, we use the LO values of $C_{1,2}$.
For all other operators, we use the NLO values in the NDR scheme,
as collected in Table~\ref{tab:wilson-scale-inputs}.
The corresponding evolution factors for the four-quark matrix elements are
\begin{equation}
 \begin{aligned}
 U_-^{\rm NLO}(\mu_H,\mu_c)&=1.16\pm0.09,\\
 U_-^{\rm NLO}(\mu_H,\sqrt{\mu_b\mu_c})&=1.29\pm0.10,\\
 U_-^{\rm NLO}(\mu_H,\mu_b)&=1.42\pm0.11.
 \end{aligned}
 \label{eq:evolution-uncertainties}
\end{equation}
{\color{black}
The uncertainties in Eq.~\eqref{eq:evolution-uncertainties} reflect the
variation of $\mu_H$ and are propagated into the lifetime uncertainties.
For the same pair of scales and to the working order, evolving the matrix
elements to $\mu_Q$ is equivalent to evolving the HQE Wilson coefficients to $\mu_H$.
The lifetime uncertainties do not include independent variation of the fixed
short-distance scales $\mu_b$, $\mu_c$, and $\mu_{bc}$.}

\section{Numerical results and discussion}
\label{sec:numerical}

For the mixing angle defined in Eq. (\ref{eq:Bbc}), the estimates based on the
quark model, QCD sum rules, and the Bethe--Salpeter equation yield a broad range of $\phi_\B$
\cite{Roberts:2007ni,Roberts:2008wq,Albertus:2009ww,Albertus:2010hi,
Aliev:2012ru,Li:2021qod}.  For the numerical analysis,
we shall use the mixing angles inferred from the bag model (see Table {\color{black}X} of
\cite{Zhang:2021yul}):
\begin{equation}
  |\Xi_{bc}(\phi_\Xi)\rangle = 0.92|1\rangle -0.39|0\rangle, \qquad
   |\Omega_{bc}(\phi_\Omega)\rangle = 0.92|1\rangle -0.40|0\rangle.
\end{equation}
{\color{black}To assess the mixing-angle sensitivity, we vary each
spectrum-derived benchmark by $\pm10^\circ$:}
\begin{equation}
 \phi_{\cal B}=
 \begin{cases}
 {\color{black}(-23.0\pm10)^\circ}, & {\cal B}=\Xi,\\
 {\color{black}(-23.5\pm10)^\circ}, & {\cal B}=\Omega .
 \end{cases}
 \label{eq:benchmark-angles}
\end{equation} 
The masses of $\Xi_{bc}$ and $\Omega_{bc}$ are given by 6.983 and 7.064 GeV, respectively \cite{Zhang:2021yul}. As for the orthogonal states,  
\begin{equation}
  |\Xi_{bc}(\phi'_\Xi)\rangle = 0.39|1\rangle +0.92|0\rangle, \qquad
   |\Omega_{bc}(\phi'_\Omega)\rangle = 0.40|1\rangle +0.92|0\rangle,
\end{equation}
their masses are 7.015 and 7.117 GeV \cite{Zhang:2021yul}, respectively, higher than that of the lower mixed $\Xi_{bc}$ and $\Omega_{bc}$ states. The higher mixed baryon states are dominated by the electromagnetic decays rather than the weak ones.

We evaluate the transition operator in
Eq.~\eqref{eq:hqe-operator-expansion} directly, using the short-distance
kernels and perturbative orders summarized in
Table~\ref{tab:coefficient-status}.  The two-quark terms combine the bilinear expansion in
Eq.~\eqref{eq:bilinear-matrix-element} with the model-specific invariant
amplitudes in Table~\ref{tab:two-quark-matrix-elements}.  Four-quark matrix
elements follow from Eq.~\eqref{eq:reduced-spin-decomposition}, using the spin
coefficients in Eq.~\eqref{eq:spin-coefficient-matrices} and
Table~\ref{tab:spin-angle-coefficients}, the reduced HBM or NRQM inputs in
Table~\ref{tab:matrix-elements}, and the flavor-dependent dimension-seven
prescription of Sec.~\ref{sec:inputs}.  The heavy--heavy contribution is
contracted directly from Eq.~\eqref{eq:tbc-tree}.  The HBM--NRQM dependence then
enters through the corresponding model matrix elements.
In the mass-sensitivity
variation, the partonic, dimension-five, Darwin, dimension-six spectator,
and dimension-seven spectator terms are scaled with their leading powers
\(m_Q^5\), \(m_Q^3\), \(m_Q^2\), \(m_Q^2\), and \(m_Q\), respectively.

\begin{figure}[!ht]
\centering
\includegraphics[width=0.98\linewidth]{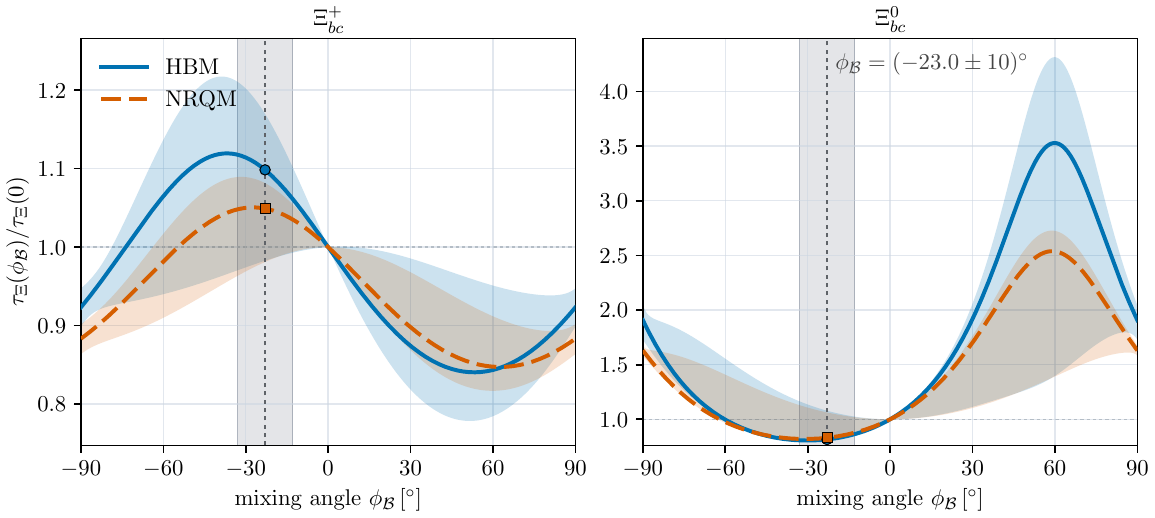}
\caption{Mixing-angle response of the
\(\Xibcp\) and \(\Xibcz\) lifetimes. {\color{black}Each curve is normalized} by
its own \(\phi_{\cal B}=0\) reference.  The parametric-sensitivity bands combine,
directionally in quadrature, the quark-model, \(m_c\), \(m_b\), and
\(\mu_H\) variations with the off-diagonal-coefficient interpolation.
The vertical dotted line marks {\color{black}the central value of}
\(\phi_{\cal B}\) in eq.~\eqref{eq:benchmark-angles}{\color{black},
with the vertical gray band showing the separate $\pm10^\circ$
interval.}}
\label{fig:xi0-mixing-angle}
\end{figure}

\begin{figure}[!ht]
\centering
\includegraphics[width=0.72\linewidth]{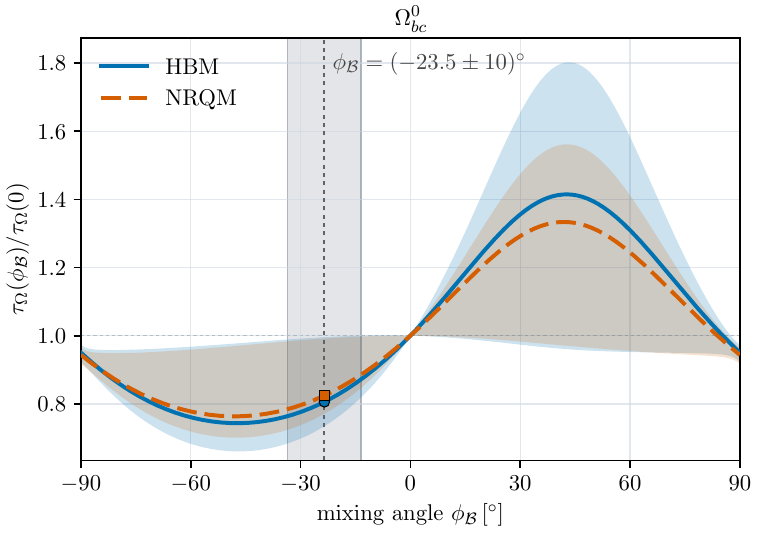}
\caption{The legend is the same as in
Fig.~\ref{fig:xi0-mixing-angle}, but the curves show the normalized
\(\Omegabcz\) lifetime.
{\color{black}The vertical gray band shows the $\pm10^\circ$ interval
around $\phi_{\cal B}=-23.5^\circ$.}}
\label{fig:omega-mixing-angle}
\end{figure}

At the flavor-specific benchmark values in
Eq.~\eqref{eq:benchmark-angles}, we obtain
\begin{align}
 \tau_{\HBM}(\Xibcp)&={\color{black}0.421^{+0.098}_{-0.093}{}^{+0.007}_{-0.014}}\,\ps,
&
 \tau_{\NRQM}(\Xibcp)&=0.367^{+0.076}_{-0.070}{\color{black}{}^{+0.001}_{-0.005}}\,\ps,
 \nonumber\\
 \tau_{\HBM}(\Xibcz)&={\color{black}0.075^{+0.035}_{-0.017}{}^{+0.004}_{-0.001}}\,\ps,
&
 \tau_{\NRQM}(\Xibcz)&=0.100^{+0.032}_{-0.014}{\color{black}{}^{+0.006}_{-0.001}}\,\ps,
 \nonumber\\
 \tau_{\HBM}(\Omegabcz)&={\color{black}0.223^{+0.074}_{-0.046}{}^{+0.018}_{-0.012}}\,\ps,
&
 \tau_{\NRQM}(\Omegabcz)&=0.229^{+0.066}_{-0.042}{\color{black}{}^{+0.017}_{-0.011}}\,\ps .
 \label{eq:benchmark-results}
\end{align}
{\color{black}The first errors are the fixed-angle uncertainties; the second
are the envelopes from varying each mixing angle by $\pm10^\circ$ with
all other inputs fixed.}
Table~\ref{tab:width-decomposition} resolves these totals as
\begin{equation}
 \Gamma_{\rm tot}
 =\Gamma_b+\Gamma_c+\Gamma_{bc}^{\rm spec},
 \qquad
 \Gamma_Q=\Gamma_{Q,2q}+\Gamma_{Q,6}^{\rm spec}
                    +\Gamma_{Q,7}^{\rm spec}\quad(Q=b,c).
 \label{eq:numerical-width-decomposition}
\end{equation}
Here \(\Gamma_b\) and \(\Gamma_c\) contain the corresponding two-quark and
heavy--light spectator terms.  The heavy--heavy
\(\Gamma_{bc}^{\rm spec}\) contraction   belongs
to one weak transition, with the \(b\) and \(c\) quarks involved
simultaneously.
Figure~\ref{fig:xi0-mixing-angle} shows the two normalized \(\Xi_{bc}\)
responses to the angles,   while 
Figure~\ref{fig:omega-mixing-angle} gives the \(\Omega_{bc}\) response.

{\it Neither} pure-spin limit is a lifetime extremum.
  Over this \(180^\circ\) period, each central
  curve has a single global minimum and maximum.  Writing each
entry as \((\phi_{\min},\phi_{\max})\), HBM and NRQM give
\((53^\circ,-37^\circ)\) and
\((63^\circ,-27^\circ)\), respectively, for \(\Xibcp\);
\((-30^\circ,60^\circ)\) and
\((-31^\circ,59^\circ)\) for \(\Xibcz\); and
\((-47^\circ,43^\circ)\) and
\((-48^\circ,42^\circ)\) for \(\Omegabcz\).
Thus an analysis restricted to the two pure configurations cannot locate
the extrema and is insufficient to characterize the full mixing-angle
dependence.
 Relative to \(\phi_{\cal B}=0\), the benchmark values reduce the neutral
\(\Xi_{bc}\) and \(\Omega_{bc}\) lifetimes by \(18\%\) and \(19\%\) in HBM,
and by \(17\%\) and \(18\%\) in NRQM, respectively, while increasing the
charged \(\Xi_{bc}\) lifetime by \(10\%\) in HBM and \(5\%\) in NRQM.
Thus the qualitative angle sensitivity survives a change of spatial wave
function despite different baselines.

\begin{table}[!ht]
\centering
\caption{Central channel decomposition.  Widths are in \(\ps^{-1}\). The brackets list
the  components of \(\Gamma_Q \).
The unshown bottom--light dimension-seven term remains included in
\(\Gamma_b\) and \(\Gamma_{\rm tot}\).   All width
columns show central values only.  The mixed-\(bc\) column is the
dimension-six heavy--heavy term.}
\label{tab:width-decomposition} 
\footnotesize
\vspace{4pt}
\setlength{\tabcolsep}{4.5pt}
\renewcommand{\arraystretch}{1.16}
\begin{tabular}{llccccc}
\toprule
Model & State
& \(\Gamma_b\,[\Gamma_{b,2q},\,\Gamma_{ b,6} ]\)
& \(\Gamma_c\,[\Gamma_{ c,2q} ,\,\Gamma_{c,6},\,\Gamma_{c,7} ]\)
& \(\Gamma_{bc}^{\rm spec}\)
& \(\Gamma_{\rm tot}\)
 \\
\midrule
\multirow{3}{*}{\(\HBM\)}
& \(\Xibcp\)
& \(0.69\,[0.65,\,+0.03]\)
& \(1.56\,[2.08,\,-1.12,\,+0.59]\)
& \(0.13\) & \(2.38\)  \\
& \(\Xibcz\)
& \(0.64\,[0.65,\,-0.01]\)
& \(12.53\,[2.08,\,+8.07,\,+2.38]\)
& \(0.13\) & \(13.31\)  \\
& \(\Omegabcz\)
& \(0.64\,[0.65,\,-0.01]\)
& \(3.73\,[2.11,\,+3.63,\,-2.01]\)
& \(0.12\) & \(4.49\)  \\
\midrule
\multirow{3}{*}{\(\NRQM\)}
& \(\Xibcp\)
& \(0.69\,[0.65,\,+0.03]\)
& \(1.82\,[2.17,\,-0.74,\,+0.40]\)
& \(0.21\) & \(2.72\)  \\
& \(\Xibcz\)
& \(0.64\,[0.65,\,-0.01]\)
& \(9.13\,[2.17,\,+5.38,\,+1.59]\)
& \(0.21\) & \(9.99\)  \\
& \(\Omegabcz\)
& \(0.64\,[0.65,\,-0.01]\)
& \(3.52\,[2.19,\,+2.67,\,-1.34]\)
& \(0.20\) & \(4.36\) \\
\bottomrule
\end{tabular}
\end{table}

The charm-initiated subtotal dominates every channel.  Constructive
\(cd\) weak exchange gives
\(\Gamma_c(\Xibcz)=12.53~\ps^{-1}\) in HBM and
\(9.13~\ps^{-1}\) in NRQM, or about \(94\%\) and \(91\%\) of the total.
Destructive \(cu\) Pauli interference instead lowers
\(\Gamma_c(\Xibcp)\) below its two-quark value, producing the longest
lifetime.  For \(\Omegabcz\),
the positive dimension-six \(cs\) term is partly cancelled at dimension
seven.  The mixed-\(bc\) term remains subleading but is enhanced in NRQM
by its larger matrix element.

\begin{table}[!ht]
\centering
\caption{Inclusive semileptonic branching fractions, in percent, at
the benchmark values of \(\phi_{\cal B}\) in
Eq.~\eqref{eq:benchmark-angles}.
{\color{black}The first errors are the fixed-angle uncertainties; the second
are the envelopes from varying each mixing angle by $\pm10^\circ$ with
all other inputs fixed.}}
\label{tab:semileptonic-branching-fractions} 
\footnotesize
\vspace{4pt} 
\setlength{\tabcolsep}{3.5pt}
\renewcommand{\arraystretch}{1.16}
\begin{tabular}{@{}llcccc@{}}
\toprule
Model & State
& \(\mathcal B_b^{\rm SL}\)
& \(\mathcal B_c^{\rm SL}\)
& \(\Delta\mathcal B_{bc}^{\rm PI,SL}\)
& \(\mathcal B_{\rm SL}^{\rm tot}\)\\
\midrule
\multirow{3}{*}{\(\HBM\)}
& \(\Xibcp\)
& {\color{black}\(6.45^{+1.56}_{-1.49}{}^{+0.11}_{-0.22}\)}
& {\color{black}\(18.63^{+2.19}_{-2.94}{}^{+0.33}_{-0.62}\)}
& {\color{black}\(+1.46^{+0.66}_{-0.59}{}^{+0.13}_{-0.12}\)}
& {\color{black}\(26.55^{+2.16}_{-3.04}{}^{+0.57}_{-0.96}\)}\\
& \(\Xibcz\)
& {\color{black}\(1.15^{+0.56}_{-0.29}{}^{+0.07}_{-0.01}\)}
& {\color{black}\(3.43^{+1.74}_{-0.97}{}^{+0.20}_{-0.04}\)}
& {\color{black}\(+0.26^{+0.09}_{-0.04}{}^{+0.02}_{-0.00}\)}
& {\color{black}\(4.84^{+2.25}_{-1.08}{}^{+0.27}_{-0.04}\)}\\
& \(\Omegabcz\)
& {\color{black}\(3.42^{+1.17}_{-0.78}{}^{+0.28}_{-0.18}\)}
& {\color{black}\(11.11^{+2.35}_{-1.52}{}^{+0.67}_{-0.43}\)}
& {\color{black}\(+0.73^{+0.26}_{-0.19}{}^{+0.02}_{-0.00}\)}
& {\color{black}\(15.26^{+3.09}_{-1.48}{}^{+0.96}_{-0.60}\)}\\
\midrule
\multirow{3}{*}{\(\NRQM\)}
& \(\Xibcp\)
& \(5.65^{+1.24}_{-1.15}{\color{black}{}^{+0.01}_{-0.07}}\)
& \(17.11^{+1.83}_{-2.31}{\color{black}{}^{+0.02}_{-0.23}}\)
& \(+2.04^{+0.48}_{-0.44}{\color{black}{}^{+0.14}_{-0.12}}\)
& \(24.81^{+0.88}_{-1.68}{\color{black}{}^{+0.12}_{-0.42}}\)\\
& \(\Xibcz\)
& \(1.54^{+0.52}_{-0.26}{\color{black}{}^{+0.09}_{-0.02}}\)
& \(4.76^{+1.57}_{-0.84}{\color{black}{}^{+0.26}_{-0.06}}\)
& \(+0.56^{+0.17}_{-0.07}{\color{black}{}^{+0.03}_{-0.00}}\)
& \(6.85^{+2.02}_{-0.66}{\color{black}{}^{+0.35}_{-0.06}}\)\\
& \(\Omegabcz\)
& \(3.53^{+1.07}_{-0.71}{\color{black}{}^{+0.26}_{-0.17}}\)
& \(13.00^{+1.53}_{-1.15}{\color{black}{}^{+0.50}_{-0.37}}\)
& \(+1.28^{+0.38}_{-0.25}{\color{black}{}^{+0.03}_{-0.00}}\)
& \(17.81^{+2.12}_{-0.70}{\color{black}{}^{+0.78}_{-0.52}}\)\\
\bottomrule
\end{tabular}
\end{table}

  We define
\(\mathcal B_b^{\rm SL}\) as the bottom-initiated two-quark
semileptonic contribution, while \(\mathcal B_c^{\rm SL}\) contains the
charm-initiated two-quark term together with any charm--light
semileptonic Pauli interference.  The signed
\(\Delta\mathcal B_{bc}^{\rm PI,SL}\) isolates the constructive
interference of the charm quark produced in \(b\to c\ell\nu\) with the
valence charm quark.  Thus the three entries sum pointwise to
\(\mathcal B_{\rm SL}^{\rm tot}\).
{\color{black}For the angular scan, the semileptonic two-quark terms are held
fixed, while the calibrated charm--light contribution follows the total
charm-spectator angular response. The second errors therefore describe
the angle sensitivity within this prescription. Numerator and denominator
are evaluated at the same angle, and the total branching fraction is
scanned directly, retaining the correlations among its components.}
The corresponding inclusive semileptonic branching fractions are
given in Table~\ref{tab:semileptonic-branching-fractions}. For a direct comparison, Ref.~\cite{Dulibic:2026} provides the central kinetic-scheme values of \(\Gamma_{\rm SL}\) and \(\Gamma_{\rm tot}\).
For the pure \(S_{bc}=0\) states, the quoted width ratios are
\(( 20, 11, 18 ) \%\) for
\((\Xi_{bc}^{+} ,  \Xi_{bc}^{0}, \Omega_{bc}^{0})\), respectively.
For the pure \(S_{bc}=1\) states, the corresponding ratios are
\((26,7,21)\%\). These branching fractions are derived from the
tabulated central widths; they are not quoted explicitly in
Ref.~\cite{Dulibic:2026}.

\begin{figure}[!th]
\centering
\hspace{-1cm}
\includegraphics[width=\linewidth]{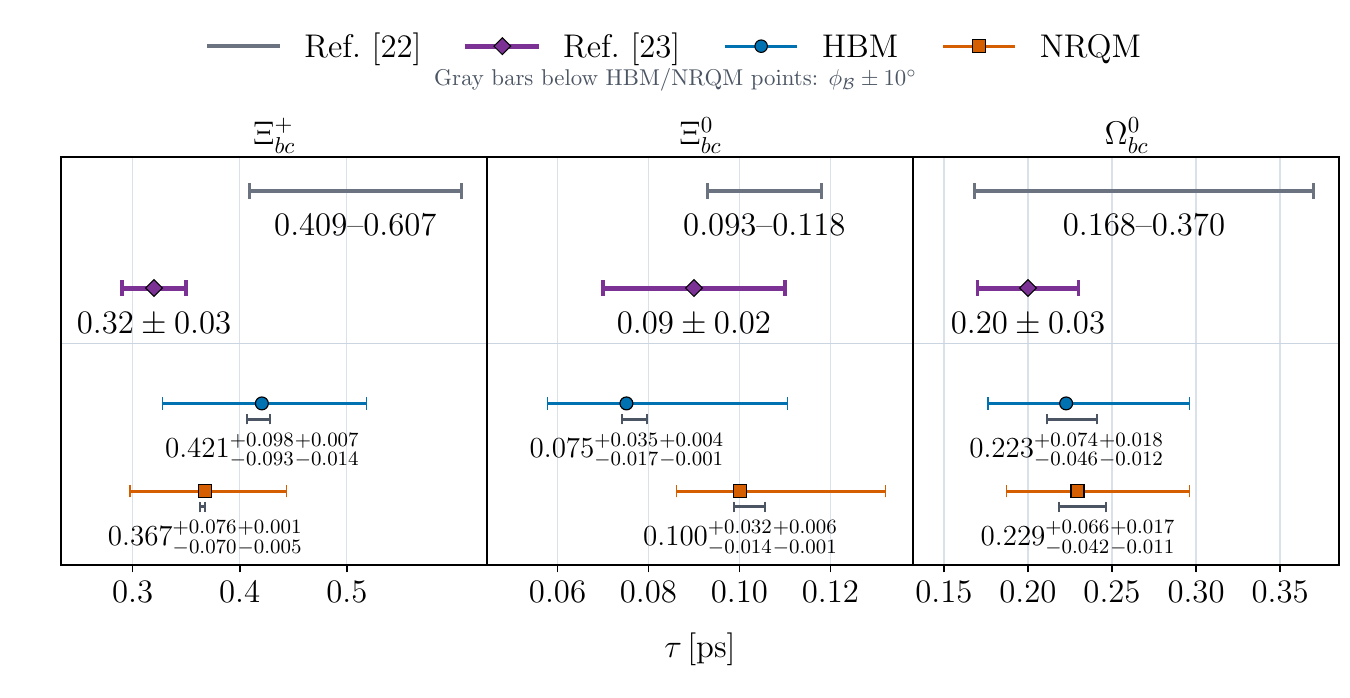}
\caption{Lifetime comparison with Ref.~\cite{Cheng:2019sxr} (gray) and
Ref.~\cite{Dulibic:2026} (purple), in \(\ps\).  The gray intervals are the
final positivity ranges; the purple intervals show the central values and
quoted parametric uncertainties.  Blue HBM and orange NRQM points show
our asymmetric parameter-variation bands at
the benchmark values of \(\phi_{\cal B}\) in
Eq.~\eqref{eq:benchmark-angles}.
Their different uncertainty prescriptions preclude a statistical
combination.}
\label{fig:three-work-comparison}
\end{figure} 

We compare our lifetime results with the literature in
Fig.~\ref{fig:three-work-comparison}.  Both
Refs.~\cite{Dulibic:2026,Cheng:2019sxr} report calculations for pure
\(S_{bc}=1\) states.  Reference~\cite{Cheng:2019sxr} uses LO coefficients
and a positivity scan of dimension-seven spectator terms, whereas
Ref.~\cite{Dulibic:2026} uses both kinetic- and $\overline{\rm MS}$-scheme coefficients and
hyperfine-based NRQM matrix elements.   Most of the difference
between our central values and Ref.~\cite{Dulibic:2026} comes from the
mixing angle.  The remaining difference reflects the
independent heavy-light and heavy-heavy matrix elements and the
heavy-light dimension-seven treatment; the comparison with
Ref.~\cite{Cheng:2019sxr} additionally involves perturbative-order and
mass-scheme differences.
 
\section{Conclusion}
\label{sec:conclusion}

We have shown that the inclusive lifetimes of
\(\Xi_{bc}\) and \(\Omega_{bc}\) cannot be characterized by the two pure
\(S_{bc}=0\) and \(S_{bc}=1\) configurations alone.  For a coherently mixed
state, the off-diagonal four-quark matrix elements generate a
\(\sin(2\phi_{\cal B})\) interference term.  Consequently, neither pure-spin
limit is a lifetime extremum: for all three baryons, the extrema of the
central curves occur at mixed configurations with
\(\lvert\phi_{\cal B}\rvert\simeq27^\circ\)--\(63^\circ\).  

We evaluated the same reduced kernels with independently calculated HBM and
Gaussian NRQM matrix elements to test whether this angular response is tied
to a particular spatial wave function.  Although the two descriptions give
different absolute overlaps and baseline lifetimes, their normalized curves
retain the same qualitative mixed-angle response.  The comparison therefore
shows that this response survives a substantial change in hadronic input;
its purpose is not to select between two model baselines.  At
\(\phi_{\Xi}=-23.0^\circ\) and \(\phi_{\Omega}=-23.5^\circ\), the
interpolation gives
 \begin{align}
 (\tau_{\Xibcp},\tau_{\Xibcz},\tau_{\Omegabcz})_{\HBM}
 &=(0.421^{+0.098}_{-0.093}{}^{+0.007}_{-0.014},\,
     0.075^{+0.035}_{-0.017}{}^{+0.004}_{-0.001},\,
     0.223^{+0.074}_{-0.046}{}^{+0.018}_{-0.012})\,\ps,\label{eq:conclusion-benchmark-hbm}\\
 (\tau_{\Xibcp},\tau_{\Xibcz},\tau_{\Omegabcz})_{\NRQM}
 &=(0.367^{+0.076}_{-0.070}{}^{+0.001}_{-0.005},\,
     0.100^{+0.032}_{-0.014}{}^{+0.006}_{-0.001},\,
     0.229^{+0.066}_{-0.042}{}^{+0.017}_{-0.011})\,\ps .
 \label{eq:conclusion-benchmark}
\end{align}
{\color{black}The first errors are the fixed-angle uncertainties. The second
errors are the separate envelopes from the $\pm10^\circ$ mixing-angle
variation in Eq.~\eqref{eq:benchmark-angles}, with all other inputs fixed.}
Relative to the corresponding spin-one references, the \(\Xibcp\) lifetime
increases by \(10\%\) in HBM and \(5\%\) in NRQM.  The \(\Xibcz\) and
\(\Omegabcz\) lifetimes decrease by \(18\%\) and \(19\%\) in HBM, and by
\(17\%\) and \(18\%\) in NRQM, respectively.  The hierarchy
\(\tau_{\Xibcp}>\tau_{\Omegabcz}>\tau_{\Xibcz}\) is controlled mainly by
charm-initiated spectator effects: destructive \(cu\) Pauli interference
makes \(\Xibcp\) the longest-lived state, constructive \(cd\) weak exchange
makes \(\Xibcz\) the shortest-lived state, and the positive dimension-six
\(cs\) contribution in \(\Omegabcz\) is partly cancelled at dimension seven.
The mixed-\(bc\) spectator term remains subleading.  The same hierarchy is
reflected in the total semileptonic branching fractions: the HBM and NRQM
central values span \(24.81\%\)--\(26.55\%\), \(4.84\%\)--\(6.85\%\), and
\(15.26\%\)--\(17.81\%\) for \(\Xibcp\), \(\Xibcz\), and
\(\Omegabcz\), respectively.

\begin{acknowledgments}
This research was supported in part by the National Science and Technology Council of the R.O.C. under Grant No. 114-2112-M-001-039, and by the National Natural Science Foundation of China under Grant Nos. 12475095 and 12575096.
\end{acknowledgments}


\begin{thebibliography}{99}

\bibitem{Shifman:1984wx}
M.~A.~Shifman and M.~B.~Voloshin,
``Preasymptotic effects in inclusive weak decays of charmed particles,''
Sov.\ J.\ Nucl.\ Phys.\ \textbf{41}, 120 (1985).

\bibitem{Shifman:1986}
M.~A.~Shifman and M.~B.~Voloshin,
``Hierarchy of lifetimes of charmed and beautiful hadrons,''
Sov.\ Phys.\ JETP \textbf{64}, 698 (1986).

\bibitem{Guberina:1986gd}
B.~Guberina, R.~Ruckl and J.~Trampetic,
``Charmed baryon lifetime differences,''
Z.\ Phys.\ C \textbf{33}, 297 (1986).

\bibitem{Chay:1990da}
J.~Chay, H.~Georgi and B.~Grinstein,
``Lepton energy distributions in heavy meson decays from QCD,''
Phys.\ Lett.\ B \textbf{247}, 399 (1990).

\bibitem{Bigi:1992su}
I.~I.~Y.~Bigi, N.~G.~Uraltsev and A.~I.~Vainshtein,
``Nonperturbative corrections to inclusive beauty and charm decays:
QCD versus phenomenological models,''
Phys.\ Lett.\ B \textbf{293}, 430 (1992);
Erratum: Phys.\ Lett.\ B \textbf{297}, 477 (1992),
arXiv:hep-ph/9207214.

\bibitem{Bigi:1993ex}
I.~I.~Y.~Bigi, M.~A.~Shifman, N.~G.~Uraltsev and A.~I.~Vainshtein,
``On the motion of heavy quarks inside hadrons: universal distributions and inclusive decays,''
Int.\ J.\ Mod.\ Phys.\ A \textbf{9}, 2467 (1994),
arXiv:hep-ph/9312359.

\bibitem{Neubert:1996we}
M.~Neubert and C.~T.~Sachrajda,
``Spectator effects in inclusive decays of beauty hadrons,''
Nucl.\ Phys.\ B \textbf{483}, 339 (1997),
arXiv:hep-ph/9603202.

\bibitem{Lenz:2014jha}
A.~Lenz,
``Lifetimes and heavy quark expansion,''
Int.\ J.\ Mod.\ Phys.\ A \textbf{30}, 1543005 (2015),
arXiv:1405.3601 [hep-ph].

\bibitem{Albrecht:2024oyn}
J.~Albrecht, F.~Bernlochner, A.~Lenz and A.~Rusov,
``Lifetimes of \(b\)-hadrons and mixing of neutral \(B\)-mesons:
theoretical and experimental status,''
Eur.\ Phys.\ J.\ ST \textbf{233}, 359 (2024),
arXiv:2402.04224 [hep-ph].


\bibitem{LHCb:2017iph}
R.~Aaij et al. [LHCb],
``Observation of the doubly charmed baryon \(\Xi_{cc}^{++}\),''
Phys.\ Rev.\ Lett.\ \textbf{119}, 112001 (2017),
arXiv:1707.01621 [hep-ex].

\bibitem{LHCb:2026xiccp}
R.~Aaij et al. [LHCb],
``Observation of the doubly charmed baryon \(\Xi_{cc}^{+}\) with the
LHCb Run~3 detector,''
Phys.\ Rev.\ Lett.\ \textbf{137}, 021902 (2026),
arXiv:2603.28456 [hep-ex].

\bibitem{YuWang} 
Y. Wang, ``Conventional spectroscopy of doubly heavy hadrons
at LHCb", talk presented at the 21st International Conference on B-Physics at Frontier Machines, Maastricht, Netherlands, June 1-5, 2026. 

\bibitem{Xu} 
A. Xu, ``Charmed-hadron properties and spectroscopy
at LHCb", talk presented at the 43rd International Conference on High Energy Physics, Natal, Brazil, July 30 to August 5, 2026. 


\bibitem{LHCb:2018zpl}
R.~Aaij et al. [LHCb],
``Measurement of the lifetime of the doubly charmed baryon \(\Xi_{cc}^{++}\),''
Phys.\ Rev.\ Lett.\ \textbf{121}, 052002 (2018),
arXiv:1806.02744 [hep-ex].

\bibitem{LHCb:2019epo}
R.~Aaij et al. [LHCb],
``Precision measurement of the \(\Xi_{cc}^{++}\) mass,''
JHEP \textbf{02}, 049 (2020),
arXiv:1911.08594 [hep-ex].



\bibitem{LHCb:2020bc0}
R.~Aaij et al. [LHCb],
``Search for the doubly heavy \(\Xi_{bc}^{0}\) baryon via decays to
\(D^0pK^-\),''
JHEP \textbf{11}, 095 (2020),
arXiv:2009.02481 [hep-ex].

\bibitem{LHCb:2021bc0}
R.~Aaij et al. [LHCb],
``Search for the doubly heavy baryons \(\Omega_{bc}^{0}\) and
\(\Xi_{bc}^{0}\) decaying to \(\Lambda_c^+\pi^-\) and \(\Xi_c^+\pi^-\),''
Chin.\ Phys.\ C \textbf{45}, 093002 (2021),
arXiv:2104.04759 [hep-ex].

\bibitem{LHCb:2022bcp}
R.~Aaij et al. [LHCb],
``Search for the doubly heavy baryon \(\Xi_{bc}^{+}\) decaying to
\(J/\psi\,\Xi_c^+\),''
Chin.\ Phys.\ C \textbf{47}, 093001 (2023),
arXiv:2204.09541 [hep-ex].

\bibitem{Roberts:2007ni}
W.~Roberts and M.~Pervin,
``Heavy baryons in a quark model,''
Int.\ J.\ Mod.\ Phys.\ A \textbf{23}, 2817 (2008),
arXiv:0711.2492 [nucl-th].

\bibitem{Li:2021qod}
Q.~Li, C.~H.~Chang, S.~X.~Qin and G.~L.~Wang,
``Mass spectra, wave functions and mixing effects of the \(bcq\) baryons,''
Eur.\ Phys.\ J.\ C \textbf{82}, 60 (2022),
arXiv:2112.10966 [hep-ph].



\bibitem{Zhang:2021yul}
W.~X.~Zhang, H.~Xu and D.~Jia,
``Masses and magnetic moments of hadrons with one and two open heavy quarks:
Heavy baryons and tetraquarks,''
Phys.\ Rev.\ D \textbf{104}, 114011 (2021),
arXiv:2109.07040 [hep-ph].


\bibitem{Cheng:2019sxr}
H.~Y.~Cheng and F.~Xu,
``Lifetimes of doubly heavy baryons \({\cal B}_{bb}\) and
\({\cal B}_{bc}\),''
Phys.\ Rev.\ D \textbf{99}, 073006 (2019),
arXiv:1903.08148 [hep-ph].


\bibitem{Dulibic:2026}
L.~Dulibi{\'c}, B.~Meli{\'c} and I.~Ni{\v{s}}and{\v{z}}i{\'c},
``New predictions for the lifetimes of doubly heavy baryons and the \(B_c\) meson,''
arXiv:2605.04967 [hep-ph].

\bibitem{Brown:2014ena}
Z.~S.~Brown, W.~Detmold, S.~Meinel and K.~Orginos,
``Charmed bottom baryon spectroscopy from lattice QCD,''
Phys.\ Rev.\ D \textbf{90}, 094507 (2014),
arXiv:1409.0497 [hep-lat].
 


\bibitem{Likhoded:1999yv}
A.~K.~Likhoded and A.~I.~Onishchenko,
``Lifetimes of doubly heavy baryons,''
arXiv:hep-ph/9912425.

\bibitem{Kiselev:1999bc}
V.~V.~Kiselev, A.~K.~Likhoded and A.~I.~Onishchenko,
``Lifetimes of \(\Xi_{bc}^{+}\) and \(\Xi_{bc}^{0}\) baryons,''
Eur.\ Phys.\ J.\ C \textbf{16}, 461 (2000),
arXiv:hep-ph/9901224.

\bibitem{Kiselev:2001fw}
V.~V.~Kiselev and A.~K.~Likhoded,
``Baryons with two heavy quarks,''
Phys.\ Usp.\ \textbf{45}, 455 (2002),
arXiv:hep-ph/0103169.


\bibitem{Karliner:2014gca}
M.~Karliner and J.~L.~Rosner,
``Baryons with two heavy quarks: masses, production, decays, and
detection,''
Phys.\ Rev.\ D \textbf{90}, 094007 (2014),
arXiv:1408.5877 [hep-ph].


\bibitem{Berezhnoy:2018}
A.~V.~Berezhnoy, A.~K.~Likhoded and A.~V.~Luchinsky,
``Doubly heavy baryons at the LHC,''
Phys.\ Rev.\ D \textbf{98}, 113004 (2018),
arXiv:1809.10058 [hep-ph].

\bibitem{Yang:2022nps}
G.~H.~Yang, E.~P.~Liang, Q.~Qin and K.~K.~Shao,
``Inclusive weak-annihilation decays and lifetimes of beauty-charmed
baryons,''
Phys.\ Rev.\ D \textbf{106}, 093013 (2022),
arXiv:2208.06834 [hep-ph].

 


\bibitem{Kiselev:1998cc}
V.~V.~Kiselev, A.~K.~Likhoded and A.~I.~Onishchenko,
``Lifetimes of doubly charmed baryons: \(\Xi_{cc}^+\) and
\(\Xi_{cc}^{++}\),''
Phys.\ Rev.\ D \textbf{60}, 014007 (1999),
arXiv:hep-ph/9807354.

\bibitem{Guberina:1999}
B.~Guberina, B.~Meli{\'c} and H.~{\v S}tefan{\v c}i{\'c},
``Inclusive decays and lifetimes of doubly charmed baryons,''
Eur.\ Phys.\ J.\ C \textbf{9}, 213 (1999);
Erratum: Eur.\ Phys.\ J.\ C \textbf{13}, 551 (2000),
arXiv:hep-ph/9901323.

\bibitem{Chang:2007xa}
C.~H.~Chang, T.~Li, X.~Q.~Li and Y.~M.~Wang,
``Lifetime of doubly charmed baryons,''
Commun.\ Theor.\ Phys.\ \textbf{49}, 993 (2008),
arXiv:0704.0016 [hep-ph].

\bibitem{Cheng:2018mwu}
H.~Y.~Cheng and Y.~L.~Shi,
``Lifetimes of doubly charmed baryons,''
Phys.\ Rev.\ D \textbf{98}, 113005 (2018),
arXiv:1809.08102 [hep-ph].


\bibitem{Dulibic:2023jeu}
L.~Dulibi{\'c}, J.~Gratrex, B.~Meli{\'c} and I.~Ni{\v{s}}and{\v{z}}i{\'c},
``Revisiting lifetimes of doubly charmed baryons,''
JHEP \textbf{07}, 061 (2023),
arXiv:2305.02243 [hep-ph].

\bibitem{Gratrex:2022xpm}
J.~Gratrex, B.~Meli{\'c} and I.~Ni{\v{s}}and{\v{z}}i{\'c},
``Lifetimes of singly charmed hadrons,''
JHEP \textbf{07}, 058 (2022),
arXiv:2204.11935 [hep-ph].

\bibitem{Gratrex:2023pfn}
J.~Gratrex, A.~Lenz, B.~Meli{\'c}, I.~Ni{\v{s}}and{\v{z}}i{\'c},
M.~L.~Piscopo and A.~V.~Rusov,
``Quark-hadron duality at work: lifetimes of bottom baryons,''
JHEP \textbf{04}, 034 (2023),
arXiv:2301.07698 [hep-ph].


\bibitem{Cheng:2023jpz}
H.~Y.~Cheng and C.~W.~Liu,
``Study of singly heavy baryon lifetimes,''
JHEP \textbf{07}, 114 (2023),
arXiv:2305.00665 [hep-ph].

\bibitem{Cheng:2026mlv}
H.~Y.~Cheng and C.~W.~Liu,
``Study of doubly heavy baryon lifetimes,''
JHEP \textbf{08}, 107 (2026),
arXiv:2604.10939 [hep-ph].



\bibitem{Hernandez:2007qv}
E.~Hernandez, J.~Nieves and J.~M.~Verde-Velasco,
``Heavy quark symmetry constraints on semileptonic form factors and decay
widths of doubly heavy baryons,''
Phys.\ Lett.\ B \textbf{663}, 234 (2008),
arXiv:0710.1186 [hep-ph].

\bibitem{Cheng:2021vca}
H.~Y.~Cheng,
``The strangest lifetime: A bizarre story of \(\tau(\Omega_c^0)\),''
Sci.\ Bull.\ \textbf{67}, 445--447 (2022),
arXiv:2111.09566 [hep-ph].

\bibitem{PDG}
F. Takahashi \textit{et al.} [Particle Data Group], Int. J. Mod. Phys. A \textbf{41}, 2630011 (2026). 

 
\bibitem{LHCb:2026OmegaBLifetime}
{\color{black}R.~Aaij et al. [LHCb],
``Measurement of the $\Omega_b^-$ baryon lifetime,''
arXiv:2609.25370 [hep-ex].}

\bibitem{Buchalla:1995vs}
G.~Buchalla, A.~J.~Buras and M.~E.~Lautenbacher,
``Weak decays beyond leading logarithms,''
Rev.\ Mod.\ Phys.\ \textbf{68}, 1125 (1996),
arXiv:hep-ph/9512380.

\bibitem{Bagan:1994zd}
E.~Bagan, P.~Ball, V.~M.~Braun and P.~Gosdzinsky,
``Charm quark mass dependence of QCD corrections to nonleptonic inclusive
\(B\) decays,''
Nucl.\ Phys.\ B \textbf{432}, 3 (1994),
arXiv:hep-ph/9408306.

\bibitem{Bagan:1994qw}
E.~Bagan, P.~Ball, V.~M.~Braun and P.~Gosdzinsky,
``Theoretical update of the semileptonic branching ratio of \(B\) mesons,''
Phys.\ Lett.\ B \textbf{342}, 362 (1995);
Erratum: Phys.\ Lett.\ B \textbf{374}, 363 (1996),
arXiv:hep-ph/9409440;
E.~Bagan, P.~Ball, B.~Fiol and P.~Gosdzinsky,
``Next-to-leading order radiative corrections to the decay
\(b\to c\bar cs\),''
Phys.\ Lett.\ B \textbf{351}, 546 (1995),
arXiv:hep-ph/9502338.

\bibitem{Krinner:2013cja}
F.~Krinner, A.~Lenz and T.~Rauh,
``The inclusive decay \(b\to c\bar cs\) revisited,''
Nucl.\ Phys.\ B \textbf{876}, 31 (2013),
arXiv:1305.5390 [hep-ph].

\bibitem{Mannel:2015jka}
T.~Mannel, A.~A.~Pivovarov and D.~Rosenthal,
``Inclusive weak decays of heavy hadrons with power suppressed terms at
NLO,''
Phys.\ Rev.\ D \textbf{92}, 054025 (2015),
arXiv:1506.08167 [hep-ph].

\bibitem{Mannel:2023zei}
T.~Mannel, D.~Moreno and A.~A.~Pivovarov,
``The heavy quark expansion for lifetimes: towards the QCD corrections to
power suppressed terms,''
Phys.\ Rev.\ D \textbf{107}, 114026 (2023),
arXiv:2304.08964 [hep-ph].

\bibitem{Lenz:2020oce}
A.~Lenz, M.~L.~Piscopo and A.~V.~Rusov,
``Contribution of the Darwin operator to non-leptonic decays of heavy
quarks,''
JHEP \textbf{12}, 199 (2020),
arXiv:2004.09527 [hep-ph].

\bibitem{King:2021xqp}
D.~King, A.~Lenz, M.~L.~Piscopo, T.~Rauh, A.~V.~Rusov and C.~Vlahos,
``Revisiting inclusive decay widths of charmed mesons,''
JHEP \textbf{08}, 241 (2022),
arXiv:2109.13219 [hep-ph].


\bibitem{Chang:2000ac}
C.~H.~Chang, S.~L.~Chen, T.~F.~Feng and X.~Q.~Li,
``The lifetime of \(B_c\) meson and some relevant problems,''
Phys.\ Rev.\ D \textbf{64}, 014003 (2001),
arXiv:hep-ph/0007162.

\bibitem{Aebischer:2021ilm}
J.~Aebischer and B.~Grinstein,
``Standard Model prediction of the \(B_c\) lifetime,''
JHEP \textbf{07}, 130 (2021),
arXiv:2105.02988 [hep-ph].

\bibitem{Lenz:2013aua}
A.~Lenz and T.~Rauh,
``\(D\)-meson lifetimes within the heavy quark expansion,''
Phys.\ Rev.\ D \textbf{88}, 034004 (2013),
arXiv:1305.3588 [hep-ph].

\bibitem{Gabbiani:2003pq}
F.~Gabbiani, A.~I.~Onishchenko and A.~A.~Petrov,
``\(\Lambda_b\) lifetime puzzle in heavy quark expansion,''
Phys.\ Rev.\ D \textbf{68}, 114006 (2003),
arXiv:hep-ph/0303235.

\bibitem{Fael:2024q2}
M.~Fael and F.~Herren,
``NNLO QCD corrections to the \(q^2\) spectrum of inclusive semileptonic
\(B\)-meson decays,''
JHEP \textbf{05}, 287 (2024),
arXiv:2403.03976 [hep-ph].

\bibitem{Egner:2024nnlo}
M.~Egner, M.~Fael, K.~Sch{\"o}nwald and M.~Steinhauser,
``Nonleptonic \(B\)-meson decays to next-to-next-to-leading order,''
JHEP \textbf{10}, 144 (2024);
Erratum: JHEP \textbf{02}, 147 (2025),
arXiv:2406.19456 [hep-ph].


\bibitem{Mannel:2024bcud}
T.~Mannel, D.~Moreno and A.~A.~Pivovarov,
``QCD corrections at subleading power for inclusive nonleptonic
\(b\to c\bar u d\) decays,''
Phys.\ Rev.\ D \textbf{110}, 094011 (2024),
arXiv:2408.06767 [hep-ph].

\bibitem{Mannel:2025bccs}
T.~Mannel, D.~Moreno and A.~A.~Pivovarov,
``QCD corrections for subleading powers in \(1/m_b\) for the nonleptonic
\(b\to c\bar c s\) transition,''
Phys.\ Rev.\ D \textbf{111}, 094035 (2025),
arXiv:2503.18775 [hep-ph].

\bibitem{Moreno:2022goo}
D.~Moreno,
``NLO QCD corrections to inclusive semitauonic weak decays of heavy
hadrons up to \(1/m_b^3\),''
Phys.\ Rev.\ D \textbf{106}, 114008 (2022),
arXiv:2207.14245 [hep-ph].

\bibitem{Moreno:2024darwin}
D.~Moreno,
``QCD corrections to the Darwin coefficient in inclusive semileptonic
\(B\to X_u\ell\bar\nu_\ell\) decays,''
Phys.\ Rev.\ D \textbf{109}, 074030 (2024),
arXiv:2402.13805 [hep-ph].

\bibitem{Falk:1994}
A.~F.~Falk, Z.~Ligeti, M.~Neubert and Y.~Nir,
``Heavy quark expansion for the inclusive decay
\(\bar B\to\tau\bar\nu X\),''
Phys.\ Lett.\ B \textbf{326}, 145 (1994),
arXiv:hep-ph/9401226.

\bibitem{Mannel:2017}
T.~Mannel, A.~V.~Rusov and F.~Shahriaran,
``Inclusive semitauonic \(B\) decays to order
\(O(\Lambda_{\rm QCD}^3/m_b^3)\),''
Nucl.\ Phys.\ B \textbf{921}, 211 (2017),
arXiv:1702.01089 [hep-ph].

\bibitem{Ciuchini:2001vx}
M.~Ciuchini, E.~Franco, V.~Lubicz and F.~Mescia,
``Next-to-leading order QCD corrections to spectator effects in lifetimes
of beauty hadrons,''
Nucl.\ Phys.\ B \textbf{625}, 211 (2002),
arXiv:hep-ph/0110375.

\bibitem{Franco:2002fc}
E.~Franco, V.~Lubicz, F.~Mescia and C.~Tarantino,
``Lifetime ratios of beauty hadrons at the next-to-leading order in QCD,''
Nucl.\ Phys.\ B \textbf{633}, 212 (2002),
arXiv:hep-ph/0203089.

\bibitem{Beneke:2002rj}
M.~Beneke, G.~Buchalla, C.~Greub, A.~Lenz and U.~Nierste,
``The \(B^+-B_d^0\) lifetime difference beyond leading logarithms,''
Nucl.\ Phys.\ B \textbf{639}, 389 (2002),
arXiv:hep-ph/0202106.

\bibitem{DeGrand:1975cf}
T.~A.~DeGrand, R.~L.~Jaffe, K.~Johnson and J.~E.~Kiskis,
``Masses and other parameters of the light hadrons,''
Phys.\ Rev.\ D \textbf{12}, 2060 (1975).

\bibitem{DeRujula:1975qlm}
A.~De Rujula, H.~Georgi and S.~L.~Glashow,
``Hadron masses in a gauge theory,''
Phys.\ Rev.\ D \textbf{12}, 147 (1975).

\bibitem{NRQM}
S.~Zeng, F.~Xu, P.~Y.~Niu and H.~Y.~Cheng,
``Doubly charmed baryon decays \(\Xi_{cc}^{++}\to\Xi_c^{(\prime)+}\pi^+\)
in the quark model,''
Phys.\ Rev.\ D \textbf{107}, 034009 (2023),
arXiv:2212.12983 [hep-ph].


\bibitem{Herren:2017osy}
F.~Herren and M.~Steinhauser,
``Version 3 of RunDec and CRunDec,''
Comput.\ Phys.\ Commun.\ \textbf{224}, 333 (2018),
arXiv:1703.03751 [hep-ph].


\bibitem{Beneke:1998ui}
M.~Beneke,
``Renormalons,''
Phys.\ Rept.\ \textbf{317}, 1 (1999),
arXiv:hep-ph/9807443.


\bibitem{Grozin:2007fh}
A.~G.~Grozin, P.~Marquard, J.~H.~Piclum and M.~Steinhauser,
``Three-loop chromomagnetic interaction in HQET,''
Nucl.\ Phys.\ B \textbf{789}, 277 (2008),
arXiv:0707.1388 [hep-ph].

\bibitem{Roberts:2008wq}
W.~Roberts and M.~Pervin,
``Hyperfine mixing and the semileptonic decays of double-heavy baryons in a
quark model,''
Int.\ J.\ Mod.\ Phys.\ A \textbf{24}, 2401 (2009),
arXiv:0803.3350 [nucl-th].

\bibitem{Albertus:2009ww}
C.~Albertus, E.~Hernandez and J.~Nieves,
``Hyperfine mixing in \(b\to c\) semileptonic decay of doubly heavy
baryons,''
Phys.\ Lett.\ B \textbf{683}, 21 (2010),
arXiv:0911.0889 [hep-ph].

\bibitem{Albertus:2010hi}
C.~Albertus, E.~Hernandez and J.~Nieves,
``Hyperfine mixing in electromagnetic decay of doubly heavy \(bc\) baryons,''
Phys.\ Lett.\ B \textbf{690}, 265 (2010),
arXiv:1004.3154 [hep-ph].

\bibitem{Aliev:2012ru}
T.~M.~Aliev, K.~Azizi and M.~Savci,
``Mixing angle of doubly heavy baryons in QCD,''
Phys.\ Lett.\ B \textbf{715}, 149 (2012),
arXiv:1205.6320 [hep-ph].

\end{thebibliography}
\end{document}